\documentclass[aps,prb.reprint,twocolumn]{revtex4} 
\usepackage{natbib}
\usepackage{epsfig}
\usepackage{graphicx}
\usepackage{amsmath}
\usepackage{amsfonts}
\usepackage{amssymb}
\usepackage{cancel}
\usepackage{graphicx}
\usepackage{wrapfig}
\usepackage[cp1251]{inputenc}
\usepackage{epsfig}

\usepackage{graphicx}
\usepackage{amsmath}
\usepackage{amsfonts}
\usepackage{amssymb}
\usepackage{graphicx}
\usepackage{wrapfig}
\usepackage{textcomp}
\usepackage{filecontents}
\usepackage{pgf,pgfarrows,pgfnodes,pgfautomata,pgfheaps,pgfshade}
\usepackage[colorlinks=true, linkcolor=blue]{hyperref}
\usepackage{xfrac}

\newcommand{{\sign}}{\rm sign}
\newcommand{\re}{\mathop{\rm Re}}
\newcommand{\im}{\mathop{\rm Im}}

\newcommand{\beq}{\begin{equation}}
\newcommand{\eeq}{\end{equation}}

\begin{document}

	\title{	Fluctuations and photon  statistics in   quantum metamaterial near the  superradiant transition 
		
	}
	
	\author{D. 
                S. Shapiro$^{1,2,3}$} \email{shapiro.dima@gmail.com}    
                \author{A.
                N. Rubtsov$^{1,3,4}$}
        \author{S. 
                 V. Remizov$^{1,2}$} 
\author{W.  
        V. Pogosov$^{1,5}$}
         \author{Yu. 
                E. Lozovik$^{6,1,7}$}  
        \affiliation{$^1$Dukhov Research Institute of Automatics (VNIIA),  Moscow 127055, Russia}
        \affiliation{$^2$V. A. Kotel'nikov Institute of Radio Engineering and Electronics, Russian Academy of Sciences, Moscow 125009, Russia}
        \affiliation{$^3$Lab of superconducting metamaterials and NTI Center for Quantum Communications, National University of Science and Technology MISiS, Moscow 119049, Russia}
        \affiliation{$^4$Russian Quantum Center,   Skolkovo, 143025 Moscow Region, Russia}
        \affiliation{$^5$Institute for Theoretical and Applied Electrodynamics, Russian Academy of
                Sciences, 125412 Moscow, Russia}
\affiliation{$^6$Institute of Spectroscopy, Russian Academy of Sciences, 142190 Moscow region,
        Troitsk, Russia}
\affiliation{$^7$Moscow Institute of Electronics and Mathematics, National Research University Higher School of Economics, 101000 Moscow, Russia}

	\begin{abstract}

The analysis of  single-mode photon  fluctuations  and their  counting statistics at the superradiant phase transition is  presented. The study concerns the equilibrium Dicke model in  a regime where the  Rabi frequency, related to a coupling of the photon mode   with a finite-number qubit environment, plays a role of the transition's control parameter. We use the  effective Matsubara action formalism  based on the representation of  Pauli operators as bilinear forms with complex and Majorana fermions.   Then,  we address    fluctuations of  superradiant order parameter and  quasiparticles. The average photon number, the fluctuational Ginzburg-Levanyuk region of the phase  transition and Fano factor are  evaluated.  We determine  the cumulant generating function which describes a full counting statistics of equilibrium photon number. Exact numerical simulation of the superradiant transition demonstrates quantitative agreement with analytical calculations.

	\end{abstract}
	\maketitle
\section{Introduction}

 The dynamics of quantum metamaterials \cite{macha2014implementation, PhysRevLett.117.210503, braumuller2017analog, Zagoskin2016, LAZARIDES20181, jung2014progress,fistul2017quantum,Shapiro2015, shapiro2015dispersive,GREENBERG2019300} 
 	is a subject of a great interest. These metamaterials are the hybrid systems where cavity photons interact with multi-qubit  environment.   The  behavior of such systems  is   captured  by   the Dicke   model \cite{emary2003chaos, brandes2005coherent, kirton2018introduction}. The interactions can be characterized by a collective Rabi  frequency  proportional to  a product of the  individual  qubit-cavity coupling constant  and square root of the qubit number.  If the Rabi  frequency    is  larger than  a certain  value
 then the superradiant phase transition, characterized by an emergence of a large  photon number in a cavity and finite order parameter, occurs for temperatures lower than a critical value.  
 The rigorous study of  the superradiant phase transition    was proposed in the  pioneering work of Fedotov and Popov \cite{popov1988functional}. These authors proposed semi-fermion parametrization of spin operators and described the phase transition in the framework of Matsubara effective action for the photon field.   In that work the chemical potential was assumed to be zero and, consequently, the  excitations' number was not constrained.  Another case of   finite chemical potential  in the Dicke model was addressed in   Refs.~\cite{eastham2001bose, eastham2006finite} and  it was shown that  the Bose condensation of polaritons is emerged   \cite{popov1988functional,eastham2006finite}
 The Keldysh diagrammatic approach for finite-$N$ corrections,  as well as effects of dissipation  and external driving, were studied in Refs.~\cite{dalla2013keldysh, PhysRevA.94.061802}.   Zero temperature description for a limit of large excitations number  
 	was obtained in Ref.~\cite{Pogosov_2017} by means of  Bethe-ansatz technique.

  Alternatively to the  temperature driven transition discussed in Ref.~\cite{popov1988functional}, the superradiance can be turned on by an  increase of the  interaction  strength. It takes place if the Rabi frequency  overcomes  a critical value.  A realization of a control parameter as the interaction energy is possible for  quantum metamaterials such as   superconducting qubits arrays \cite{macha2014implementation, PhysRevLett.117.210503, Shulga2017, Zhang2017}   integrated with a GHz transmission line via tunable couplers \cite{Srinivasan2011,Hoffman2011,Chen2014,Zeytinouglu2015}. Also, this may be done in hybrid systems with a controllable amount of nitrogen-vacancy (NV) centers in a diamond sample which  interact with an electromagnetic field    \cite{Dutt2007,Sandner2012,Putz2014,Angerer2018}.

  In the present paper we address  the situation where the Rabi frequency in  quantum metamaterial is varied from weak to ultra-strong coupling domains  while the temperature remains  constant. We also keep a constant number of qubits $N$  assuming that $N $ is large but finite.  It is implied  that  the loss rate in the cavity is   small.   The finiteness of $N$ in our consideration means that the  superradiant transition is smoothed by the fluctuations of the order parameter and, beside of that, by the thermal fluctuations of  polariton quasiparticles.  The aim of this work is (i) to describe fluctuations of the above two types and (ii) to formulate a full counting statistics for the  photon numbers  in this regime.

 Our main results are the   explicit expressions for the  average photons number, its fluctuations and full counting statistics 
     as   functions of the collective Rabi frequency.  Proposed  formalism provides a solution for  low temperature    $T$ and large   $N$ provided they satisfy the  conditions  $\hbar\omega\gg k_{\rm B}T\gg \hbar\omega/N$ (in this case all qubits are assumed to be in a resonance with  the photon  mode of the  frequency $\omega$). The generalizations for the high-temperature limit, $k_{\rm B}T\gg\hbar\omega$, and dispersive regime, where a spectral density of qubits energies is strongly broadened, are also discussed.

The paper is organized as follows.
In the Sec. \ref{sec:path_int} we present Matsubara action for the Dicke model, where the  qubits degrees of freedom are expressed through the Majorana and complex fermion  variables. This is one of possible representations of Pauli operators acting in a Hilbert space of a two-level system.   The Sec. \ref{sec:eff_action} has methodological character. We derive  the photon mode's effective action, which was obtained in previous works~\cite{popov1988functional,eastham2006finite}, by means of the  alternative technique with Majorana fermions.   In the Sec.  \ref{sec:n_ph} we present the  general expressions for the average photon number and their fluctuations   in a resonant limit.  In the Sec.  \ref{sec:ph_tr}    we discuss  fluctuational and statistical properties  
and present a comparison with  results of exact numerical simulations at finite temperature and qubit number of the order of ten. 
In Sec. \ref{sec:generalization}  the results  are generalized for high temperatures and inhomogeneous broadening in qubit ensemble. In Sec. \ref{sec:fcs}  the   cumulant generating function for the photon number is derived. In the Sec. \ref{sec:concl} 
we conclude. In the Appendix \ref{app-corr} we 
derive the  conditions, where our solution based on the Gaussian approximation for thermal  fluctuations is strict.

	 \section{Path integral formulation}\label{sec:path_int}

The Dicke  Hamiltonian  of $N$ qubits reads (we set $\hbar=1$ and $k_{\rm B}=1$ throughout the text):
 \begin{equation}
\hat H =\omega\hat \psi^\dagger\hat \psi+\sum\limits_{j=1}^N \frac{\epsilon_j}{2}\hat\sigma^z_j+  \sum\limits_{j=1}^N   g_j (\hat\psi\hat\sigma^+_j  + \hat\psi^\dagger\sigma^-_j).
\label{h-rwa}
\end{equation}
Here $\epsilon_j$ are the qubits excitation energies, $g_j$ are the individual  coupling strengths between $j$-th qubit and the photon field in a single-mode cavity. The fundamental frequency of the photon mode is $\omega$. The coupling term is introduced in the standard rotating wave approximation. 

In a path integral formulation   the photon mode is described by a conventional complex bosonic fields $\bar\psi,\psi$.
The   Pauli operators, $\hat \sigma^\pm_j,\hat \sigma^z_j$,  acting on the $j$-th qubit degrees of freedom,  may be represented in path integrals in different ways.
 It can be  bosonic Holstein-Primakoff  representation \cite{PhysRev.58.1098}  or   bilinear forms of fermions. 
 Concerning other  fermion representations for the Dicke model,  techniques based on semi-fermions with an  imaginary chemical potential \cite{popov1988functional} or   auxiliary  boson field \cite{eastham2001bose} were employed. These representations allow  to  eliminate the emergent unphysical states  and to reduce a Hilbert space to that of a spin-1/2.  The semi-fermion representation for spin operators was generalized for Keldysh technique  in Ref.~\cite{PhysRevLett.85.5631}.
  Another one fermion representation, which we choose for our calculations,  is given by the product of a complex $\hat c_j\neq \hat c^\dagger_j$ and Majorana $\hat d_j=\hat d_j^\dagger$ fermion operators~\cite{martin1959generalized,tsvelik2007quantum}:
\begin{equation}
\hat\sigma^+_j=\sqrt{2}\hat c^\dagger_j \hat d_j, \quad \hat \sigma^-_j=\sqrt{2}\hat d_j\hat c_j  . \label{fermionization}
\end{equation}  
They correspond  to three Grassmann fields   $\bar c ,c$ and $d$ in a path integral formalism. The use of Majorana fermion allows to avoid  auxiliary constraints in the action.  Fields $\bar c, c$  are related to usual complex fermion mode with  the excitation energy of  two-level system.  Field $d$ stands for Majorana zero energy mode with  $\langle \hat d^2\rangle=1/2$.   Majorana representation of spin operators has been   
	applied to spin-boson model \cite{SCHAD2015401,PhysRevB.93.174420} and to a description  of spin-spin interaction via helical Luttinger liquid  \cite{PhysRevLett.120.147201}. Recently, this fermionization has been applied to the Dicke model with counter-rotating terms in the interaction Hamiltonian and  a regime of quantum chaos  has been studied   \cite{1808.02038v2}. In our studies, which are focused on    fluctuation-dominated regime  near superradiant phase transition and behavior at  finite $N$, Majorana  representation appears  as  a convenient tool.

Below we demonstrate  how one can obtain the effective action for photon field  with the use of the fermionization  (\ref{fermionization}).
 The starting point of such consideration is the path integral formulation of the  partition function $Z$  in terms of the  boson complex fields $\Psi_\tau=[\bar\psi_\tau,\psi_\tau]$ and fermion fields $\bar c ,c,d$~\cite{kamenev2011field}:
\begin{equation}
 Z=
 	\int \mathcal{D}[\Psi, \bar c ,c,d ]
 	\exp(- S[\Psi, \bar c ,c,d ]) 
\label{Z}
\end{equation}
with the action is 
\begin{multline}S [\Psi, \bar c ,c,d ]=S_{\rm ph}[\Psi ] +  S_{\rm q} [  \bar c ,c,d ] + \\ + S_{\rm int}[\Psi, \bar c ,c,d ]+\ln Z_{\rm ph}Z_{\rm q}  \ .  \label{S} \end{multline}
Here $S_{\rm ph}[\Psi ] $, $S_{\rm q} [  \bar c ,c,d ] $ and $S_{\rm int}[\Psi, \bar c ,c,d ]$ are the Matsubara actions of the photon mode, qubit environment and their interaction, respectively. The last term $\ln Z_{\rm ph}Z_{\rm q}$  appears  due to a normalization of $Z$  to unity at  the  decoupled limit  $g_j \to 0$.

 Below we consider the  terms in (\ref{S}) in more details.  Both of the qubit and photon subsystems are assumed to be in thermal equilibrium at the temperature $T$. The   photon mode action, defined on the imaginary time interval $\tau\in [0,\beta]$, where $\beta=1/T$, is 
\begin{equation}
	S_{\rm ph}[\Psi]  =\int\limits_0^\beta  \bar \psi_\tau(-G_{{\rm ph}; \tau-\tau'})\psi_{\tau'} \ d\tau  , 
	\label{Sph}
\end{equation}
where the inverse Green function of free photon mode   is 
\begin{equation}
	\quad G_{{\rm ph}; \tau-\tau'}^{-1}=\delta_{\tau-\tau'}(-\partial_{\tau'}-\omega)
	\ .
	\end{equation}
The Fourier transformations from $\tau $ to Matsubara bosonic frequencies  $\omega_n=2\pi n T$ are defined  for the fields 
 and for the Green functions as 
\begin{equation}
	\psi_n=T \int\limits_0^\beta \psi_\tau e^{{\rm i}2\pi n T \tau} d\tau, \ \bar\psi_n=T \int\limits_0^\beta \bar\psi_\tau e^{-{\rm i}2\pi n T \tau} d\tau
\end{equation}
and 
 \begin{equation}
 	G_{{\rm ph}; n}^{-1}= \int\limits_0^\beta G_{{\rm ph};\tau }^{-1} e^{{\rm i}2\pi n T \tau} d\tau={\rm i}2\pi n T -\omega. 
 \end{equation}
In this representation   the photon mode action (\ref{Sph})   is transformed into  
 \begin{equation}
 	S_{\rm ph}[\Psi]  =\beta\sum\limits_{n }\bar\psi_n (-G_{{\rm ph}; n}^{-1} ) \psi_n .
 	\label{Sph-1}
 \end{equation} 
The qubit ensemble action is
\begin{multline}
	S_{\rm q}[ \bar c ,c,d ] \\
	=\frac{1}{2}\sum\limits_{j=1}^N\int\limits_0^\beta \begin{bmatrix} \bar c_j & c_j & d_j  
	\end{bmatrix}
	(-\mathbf{G}_{j;\tau-\tau'}^{-1} )
	\begin{bmatrix}  c_j \\   \bar c_j  \\  d_j  
	\end{bmatrix}  d\tau   d\tau'  \ .
	\label{Sq}
\end{multline}
The matrix $\mathbf{G}_{j;\tau-\tau'}^{-1}$ describes the $j$-th qubit. It contains the inverse Green functions for the $j$th complex fermion and its conjugate with the  energies $\pm \epsilon_j$, respectively, and the Majorana fermion of zero energy:
\begin{equation}
	 -  \mathbf{G}_{j;\tau-\tau'}^{-1}=\delta_{\tau-\tau'}
\begin{bmatrix} \partial_{\tau'} + \epsilon_j  && 0 && 0 \\ \\
	0 && 	\partial_{\tau'} -\epsilon_j   && 0 \\ \\
	0  && 0 && \partial_{\tau'} 
\end{bmatrix} \ .\label{G}
	\end{equation}
Note, that a corresponding Fourier transformation of the fields $\bar c_\tau, c_\tau$ and $d_\tau$ and the elements of $\mathbf{G}_{j;\tau-\tau'}$ assumes the  fermionic frequencies $\omega_n=2\pi n T+\pi T$. Bilinear forms $c_j d_j$ and $\bar c_j d_j$  appear in $S [\Psi, \bar c ,c,d ]$ due to the qubit-cavity coupling encoded by the matrix  $\mathbf{V}_j[\Psi_\tau]$:
\begin{multline}
	S_{\rm int}[\Psi, \bar c ,c,d ]\\ =
	\frac{1}{2}\sum\limits_{j=1}^N\int\limits_0^\beta \begin{bmatrix} \bar c_j & c_j & d_j  
	\end{bmatrix}
	\delta_{\tau-\tau'}\mathbf{V}_j [\Psi_\tau]
	\begin{bmatrix}  c_j \\   \bar c_j  \\  d_j  
	\end{bmatrix}  d\tau  d\tau' .
	\label{Sint}
\end{multline}
This is the  matrix which involves the complex boson fields  $\psi_{\tau}$, $\bar\psi_{\tau}$  as follows:
\begin{equation}
	\mathbf{V}_j[\Psi_\tau] =\sqrt{2}g_j 
	\begin{bmatrix} 0  && 0 && -\psi_\tau \\ \\
	0 && 	0   &&  \bar\psi_\tau \\ \\
	-\bar\psi_\tau  && \psi_\tau && 0
	\end{bmatrix}.\label{V}
\end{equation}
The normalization term in (\ref{S}) is the product of partition functions of    non-interacting  photon mode  and $N$ qubits.
The logarithms of their partition functions $ Z_{\rm ph}=\int \mathcal{D}[\Psi ]  \exp(-S_{\rm ph}[\Psi]  )$ and  $Z_{\rm q}=\int \mathcal{D}[ \bar c ,c,d ]  \exp(  - S_{\rm q}[\bar c ,c,d ])$ are the following:
\begin{equation}
	\ln Z_{{\rm ph} } =- {\rm Tr}\ln (- G ^{-1}_{{\rm ph};\tau-\tau'}) 
\end{equation}
and 
\begin{equation}
	\ln Z_{{\rm q} } =\frac{1}{2}\sum\limits_{j=1}^{N}{\rm Tr} \ln (-\mathbf{G}^{-1}_{j;\tau-\tau'}) \ .
\end{equation}
The prefactor of \sfrac{1}{2}
results from   the integration over Grassmann variables in the representation (\ref{Sq}). 
The sign  ``${\rm Tr}$''  means the trace taken over the imaginary time variables, or, equivalently, by the  Matsubara frequency  index $n$;  in a case of qubits, an additional trace is taken over the internal $3\times 3$  structure of a matrix $\mathbf{G} _{j}$.

\section{Effective action}\label{sec:eff_action} 

 To derive  the effective action for the photon field, $S_{\rm eff}[\Psi]$,  from the full one $S [\Psi, \bar c ,c,d ]$, we start from  integration over the the fermion modes $c_j, \bar c_j$ and $d_j$.
As a result, the path integral in the  partition function is reduced to   $ 
Z=\int D[ \Psi] e^{-S_{\rm eff}[\Psi]} 
$
where the effective action is obtained in the most general  form  
\begin{multline}
	 S_{\rm eff}[\Psi]=S_{\rm ph}[\Psi] +\ln Z_{\rm ph}Z_{\rm q}-\\ -\frac{1}{2}
	 {\rm Tr} \ln (-\mathbf{G}^{-1}_{j;\tau-\tau'} +\delta_{\tau-\tau'} \mathbf{V}_j[\Psi_\tau]). \label{s_eff_1}
\end{multline}
Expanding the logarithm in the last term of (\ref{s_eff_1}) we obtain that  all  odd order terms are equal to zero. This  follows from the diagonal  and non-diagonal structures of $\mathbf{G}_j$ and  $\mathbf{V}_j$, respectively. The resummation back of the non-zero terms of even orders   gives the identity:
\begin{multline}
{\rm Tr} \ln (-\mathbf{G}^{-1}_{j;\tau-\tau'} +\delta_{\tau-\tau'} \mathbf{V}_j[\Psi_\tau])=\ln Z_{{\rm q} }+\\
+\frac{1}{2}{\rm Tr} \ln (-\mathbf{G}^{-1}_{j;\tau-\tau'} + \mathbf{V}_j[\Psi_\tau]\mathbf{G}_{j;\tau-\tau'} \mathbf{V}_j[\Psi_{\tau'}]) \ . \label{tr_log}
\end{multline}

A direct first order expansion of the logarithm in the second line of (\ref{tr_log})	 by $\mathbf{V}[\Psi_\tau] \mathbf{G}_{\tau-\tau'} \mathbf{V}[\Psi_{\tau'}]$  provides  Gaussian action for all Matsubara modes  $\bar\psi_n$, $\psi_n$. As it will be shown in Sec. \ref{sec:n_ph},  this expansion results in divergent number of photons at the critical Rabi  frequency  near the transition into superradiant phase  (see Eq.~\ref{NphGauss}). This follows from  an infinite occupation of  zeroth Matsubara frequency  component of the field  \begin{equation}
	\psi_0\equiv T\int\limits_0^\beta\psi_\tau d\tau \ .
	\end{equation}
To make correct description of photonic subsystem we should leave $\psi_0$ in zero order term of (\ref{tr_log})  and expand the logarithm by the fluctuations $\delta\psi_\tau\equiv \psi_\tau- \psi_0$. This results in effective regularization of the divergency.
Note, that Fourier transformation $\delta\psi_\tau$  gives the non-zero Matsubara components $\psi_{n\neq 0}$.
The field $\psi_0$ is  related to the complex amplitude of a superradiant order parameter while   $\psi_{n\neq 0}$ are related to thermal fluctuations of polaritonic quasiparticles.

The regularization of the divergency mentioned above  assumes a  redefinition of the Green function,  $\mathbf{G}_{j}\to \mathcal{G}_j[\Psi_0]$  with $ \Psi_0=[\bar\psi_0,\psi_0]$, as follows:
 \begin{equation}
 	\mathcal{G}^{-1}_{j;\tau-\tau'}[\Psi_0]\equiv \mathbf{G}^{-1}_{j;\tau-\tau'} -   \mathbf{V}_j[\Psi_0]\mathbf{G}_{j;\tau-\tau'} \mathbf{V}_j[\Psi_0] \ .
 \end{equation}
Here we introduce   the matrix with zero-mode components
 \begin{equation}
\mathbf{V}_j[\Psi_0]=\frac{1}{\beta}\int\limits_0^\beta \mathbf{V}_j[\Psi_\tau]  d\tau \ .
 \end{equation}

  Below we limit our consideration of the fluctuations taking into account  bilinear combinations of the fields $\delta\bar\psi_\tau $ and $\delta\psi_{\tau'}$.  These are   gauge invariant terms $\delta\bar\psi_\tau\delta\psi_{\tau'}$ which provide normal coupling channel between  the photons in the dissipative action $S_\Sigma$. Oppositely, terms   $\delta\bar\psi_\tau\bar\delta\psi_{\tau'}$ and $ \delta\psi_\tau\delta\psi_{\tau'}$ are not gauge invariant and   provide      anomalous type of   coupling.  
 At the given  step of the derivation we perform the logarithm expansion   in  (\ref{tr_log}) around $\mathcal{G}^{-1}$  in second order by the matrix
\begin{equation}
\mathbf{V}_j[\delta\Psi_\tau]\equiv\mathbf{V}_j[\Psi_\tau] - \mathbf{V}[\Psi_0] \ ,
\end{equation}
 which involves the  fluctuating parts in  $\delta\Psi_\tau=[\delta\bar\psi_\tau,\delta\psi_\tau]$. We note that the first order contribution by $\mathbf{V}_j[\delta\Psi_\tau]$ equals zero in this approach.
 As a result, we obtain 
\begin{equation}
	S_{\rm eff}[\Psi]=S_{\rm ph}[\Psi]+S_{\mathcal{G}}[\Psi_0]+S_{\rm \Sigma}[\Psi]+\ln Z_{\rm ph} \ . 
	\label{s_eff}
\end{equation}
The first term $S_{\rm ph}$ in (\ref{s_eff}) is not changed. The second term $S_{\mathcal{G}}[\Psi_0]\equiv -\frac{1}{4}\sum_j{\rm Tr} \ln\left( \mathbf{G}_j\mathcal{G}_j^{-1}[\Psi_0]\right)$  involves the zero-frequency mode  $\Psi_0$ only.  Note, that in the Dicke model (\ref{h-rwa})  the interaction   is limited by the  rotating wave approximation which   conserves the excitations number. In this case $S_{\mathcal{G}}$ depends on the zero mode's magnitude squared, $\Phi\equiv \bar\psi_0\psi_0$,  and is independent on its complex phase $\varphi \equiv \arg \psi_0$. Thus, $S_{\mathcal{G}}[\Psi_0]= S_{\mathcal{G}}[\Phi] $   and its explicit expression is 
\begin{equation}
	S_{\mathcal{G}}[\Phi]=  
	-  \sum\limits_{j=1}^N \ln \frac{\cosh\frac{\sqrt{\epsilon_j^2+4g_j^2 \Phi 
	}}{2T}}{\cosh\frac{ \epsilon_j } {2T}} \ . \label{s-zm}
\end{equation}
 This result follows from a representation of the Green functions $\mathbf{G}$ and $\mathcal{G}$ in Matsubara frequencies $\omega_n$. After that,   $S_{\mathcal{G}}$ is reduced to a calculation of infinite product by $n$.  
  The third term in (\ref{s_eff})  quadratic by quasiparticle  fluctuations reads 
\begin{multline}S_{\Sigma}[\Psi]=\\
	\frac{\beta}{2}\sum\limits_{n\neq 0}\begin{bmatrix}  \bar\psi_n &   \psi_{-n}
	\end{bmatrix}
	\begin{bmatrix}
		\Sigma_n[\Psi_0] && \tilde\Sigma_n[\Psi_0] \\  \\
		(\tilde\Sigma_{-n}[\Psi_0])^* && \tilde\Sigma_{-n}[\Psi_0]
	\end{bmatrix}
	\begin{bmatrix}  \psi_n \\ \\   \bar\psi_{-n}
	\end{bmatrix} \ . \label{s-fl} 
\end{multline}
This is the  dissipative part of the action; it corresponds to  effective photon-photon interaction via qubits degrees of freedom.  
  The self-energy operators $\Sigma_{\tau }[ \Psi_0 ]$ and $\tilde\Sigma_{\tau }[ \Psi_0 ]$ provide normal and anomalous channels of the photon-photon interactions, respectively. 
  They  result  from a summation over the fermionic Matsubara frequencies.
From calculations it follows  that normal self-energy depends on $\Phi$ only, $\Sigma[\Psi_0]=\Sigma[\Phi]$, while the anomalous one depends also on the phase, i.e., $\tilde\Sigma[\Psi_0]\equiv \tilde\Sigma[\Phi,\varphi]$. Their explicit expressions are presented in the Appendix \ref{app-corr}, see Eqs.~(\ref{sigma-normal}) and (\ref{sigma-anomal}). The above results for $S_{\mathcal{G}}$ and $S_{\rm \Sigma}$ are in full correspondence with that derived in Refs.~\cite{popov1988functional,eastham2006finite} using alternative spin representations.

 The action $S_{\rm eff}$ allows to calculate the thermodynamical average value $\langle\Phi\rangle$ which is   superradiant order parameter.  As shown below,  the action indicates the superradiance as a second order phase transition. The quadratic expansion by $\Phi$ in 
 $S_{\mathcal{G}}[\Phi]$
 allows to capture this transition (it corresponds to taking into account the  non-Gaussian   $|\psi_0|^4$).
   As a consequence, if the system is in the normal phase or near the phase transition,  one  can simplify $S_{\mathcal{G}}$ and $S_{\rm \Sigma}$ assuming that the relevant values of $\Phi$ belongs to a certain region near $\Phi=0$.   Namely, analytical calculations presented in this work assumes that we apply second order expansion by $\Phi$ in $S_{\mathcal{G}}$ and neglect  by  non-Gaussian cross terms $\propto  \Phi\bar\psi_n\psi_n $ in  $S_{\rm \Sigma}$. For the zero-mode part it means
	\begin{equation}S_{\mathcal{G}}[\Phi]\approx\Phi S_{\mathcal{G}}'[0]+\frac{1}{2}\Phi^2 S_{\mathcal{G}}''[0] \ . \label{s_zm_quadratic}
	\end{equation}
For $S_{\rm \Sigma}$ part it means one can neglect   the dependencies of self-energies  on $\Phi$ and   assume 
\begin{equation}
	S_{\rm \Sigma}[\Psi]\approx S_{\rm \Sigma}[\Phi{=}0,\delta\Psi] \ . \label{s-fl-simplif}
\end{equation}
We obtain that this approximation involves the normal coupling only, i.e., 
\begin{equation}
	 S_{\rm \Sigma}[\Phi{=}0,\delta\Psi] = \beta\sum\limits_{n\neq 0} \Sigma_n[0] \bar\psi_n\psi_n \ . \label{s-fl-simplif-1}
\end{equation}
It follows from (\ref{sigma-anomal})  that $
\tilde\Sigma_n[\Phi,\varphi]\propto \Phi \ 
$ for small $\Phi$ 
and, hence,  the anomalous terms does not appear in (\ref{s-fl-simplif}). Note, that  $S_{\rm \Sigma}$ is purely Gaussian in this case because the terms proportional to $\Phi \bar\psi_n\psi_n$ are neglected.

 The validity of the approximations (\ref{s_zm_quadratic}) and (\ref{s-fl-simplif}) is analyzed in the Appendix \ref{app-corr} by means of the effective action for the zero Matsubara mode, see Eq. (\ref{app:s-0}). This action is obtained after the  Gaussian integration over all non-zero modes  $\psi_{n\neq 0}$ in $S_{\rm eff}$ from   (\ref{s_eff}).  The linear by $\Phi$ contributions to the self-energies, $\Sigma_n[ \Phi ]\approx\Sigma_n[ 0 ]+\Phi\Sigma'_n[ 0 ] $ and $\tilde\Sigma_n[\Phi,\varphi]\propto \Phi$, are investigated as a perturbations for the action (\ref{app:s-0}). 
It is shown   that such perturbations are small and  approximations (\ref{s_zm_quadratic}) and (\ref{s-fl-simplif}) are strict    if 
 the condition for the temperature and qubit number
\begin{equation}
	T\gg \frac{\omega}{N} \label{condition-0}
\end{equation}
holds. It is assumed here that  qubits' and photon mode's frequencies are of the same order, $\epsilon_j\sim \omega$. 
 The condition (\ref{condition-0}) also provides the range of parameters where one can go beyond the thermodynamic limit and study finite-$N$ effects. 
  The thermodynamic limit, where fluctuations   of order parameter are negligible as $1/N$, corresponds to a situation of simultaneous limits $N\to \infty$ and $g\to 0$ with the constraint $\sqrt{N}g={\rm const}$.

  As it is also shown in Appendix \ref{app-corr},  the ratio \begin{equation}\kappa_{\rm c}=\sqrt{\frac{\omega}{NT}}\ll 1 \label{kappa}
	\end{equation} provides the small parameter of this theory near the superradiant transition which allows one to neglect   non-Gaussian terms in a controllable way. 
 
  To summarize the above, for large enough qubit number dictated by (\ref{condition-0}), we obtain an effective theory for low temperatures,  $\omega\gg T\gg \omega/N$. The  high  temperature domain corresponds to  $T\gg\omega$. Two  approximations (\ref{s-fl-simplif}) and (\ref{s_zm_quadratic}) yield the effective action $S_{\rm eff,0}$ which  provides a description of  the  normal phase and fluctuational region near the superradiant transition.    
 It is convenient to represent it as   
\begin{equation}
S_{\rm eff,0}[\Phi,\bar\psi_n,\psi_n]=S_{\rm zm}[\Phi]+S_{\rm fl}[\bar\psi_n,\psi_n]+\ln Z_{\rm ph} \ .	\label{s_eff0}
	\end{equation}
where the zero-mode terms are collected in
\begin{equation}
S_{\rm zm}[\Phi]=A\Phi+{\it \Gamma} \Phi^2 \ 
	\end{equation}
and that of quasiparticle fluctuations in 
\begin{equation}
S_{\rm fl}[\bar\psi_n,\psi_n]=\beta\sum\limits_{n\neq 0}(-{\rm i}2\pi n T+\omega+\Sigma_n[0])\bar\psi_n\psi_n \ . \label{s-fl-main}
	\end{equation}
The parameters for a general case are:
\begin{equation}
	A=\beta \omega- \beta \sum\limits_{j=1}^N\frac{g^2_j}{\epsilon_j}\tanh\frac{\beta\epsilon_j}{2} \ , 
	\label{alpha}
\end{equation}
\begin{equation}
	{\it \Gamma}=  \beta \sum\limits_{j=1}^Ng^4_j\frac{\sinh \beta\epsilon_j - \beta\epsilon_j}{\epsilon_j^3(\cosh\beta\epsilon_j +1)}, \label{gamma}
\end{equation}
and 
\begin{equation}
	\Sigma_n[ 0 ]= \sum\limits_{j=1}^N  \frac{g^2_j\tanh\frac{\epsilon_j}{2T}}{2 {\rm i}\pi n T -  \epsilon_j}. \label{sigma-0}
\end{equation}
In the above formulation, the critical point is $A=0$. 
For $A>0$ the system is in the normal phase and for $A<0$ a superradiant phase with large amount of photons does emerge.   In other words, if $A<0$ then   $S_{\rm eff,0}$ has  a minimum   at  the stationary point $\Phi=\Phi^*$  with  \begin{equation}\Phi^*=-\frac{A}{2{\it \Gamma}} \ .\label{saddle-point-0}
	\end{equation}
	In terms of the complex photon field  $\psi_0$ this corresponds to a saddle line which is a circle in the complex plane.

The control parameter of the phase transition is the collective Rabi frequency defined as
\begin{equation}
	\Omega=\sqrt{N\langle g^2\rangle_j}
     ,\qquad 
        \langle g^2\rangle_j = \frac{1}{N}\sum_{j=1}^N g_j^2
       \  , 
\end{equation}
 where we denote $\langle \cdot \rangle_j$ as the average  over the qubit ensemble. The superradiance condition $A<0$ corresponds to the Rabi frequency  exceeding a certain critical value $\Omega>\Omega_{\rm c}$.
For the homogeneous limit where all qubits have the same energy, $\epsilon_j=\bar\epsilon$, the saddle point (\ref{saddle-point-0}) is given by 
\begin{equation}
	\Phi^*=N\left(\frac{\bar\epsilon}{\Omega}\right)^2\left(\tanh\frac{\bar\epsilon }{2T}- \frac{\bar\epsilon\omega}{\Omega^2}\right)  \frac{1+\cosh\beta\bar\epsilon}{\sinh\beta\bar\epsilon-\beta\bar\epsilon}  . \label{saddle-point}
\end{equation} 
The  critical Rabi frequency of the phase transition follows from the condition $\Phi^*=0$. From (\ref{saddle-point}) one finds that
\begin{equation} 
\Omega_{\rm c}=  \sqrt{\bar\epsilon\omega \coth\frac{ \bar\epsilon   }{2T}}.  \label{omega-c}
\end{equation}

We also introduce the action 
 \begin{equation}S_{\rm eff,1}[\Psi]=    S_{\rm ph}[\Psi]+S_{\rm fl}[\bar\psi_n,\psi_n]+ S_{\mathcal{G}} [\Phi]    \label{s_eff1}
\end{equation}
 where, in contrast to $S_{\rm eff,0}[\Psi]$ in (\ref{s_eff0}), the zero mode's part (\ref{s-zm}) is taken into account exactly and  its logarithm  is not expanded.
This action provides an adequate description of the superradiant phase where is $\langle\Phi\rangle$ large. Calculations combine the exact integration by $\bar\psi_n$ and $\psi_n$ and numerical integration by $\Phi$ in this regime. In what follows we focus mainly on the transition between the normal phase and the  fluctuational region  employing the formalism of  $S_{\rm eff,0}[\Psi]$.   A behavior in the superradiant phase is briefly discussed  below.

\section{Photons number and their fluctuations for resonant limit}\label{sec:n_ph}
In this part of the paper 
 we study fluctuational behavior of the superradiance with the use of $S_{\rm eff,0}$ in a limit of full resonance between qubits and photon mode, i.e., 
 \begin{equation}
\epsilon_j=\bar\epsilon=\omega. \label{res}
\end{equation}
The disorder in $g_j$, in its turn, is taken into account. The   parameters (\ref{alpha}, \ref{gamma}) are reduced to 
\begin{equation}
 \alpha\equiv A_{\epsilon_j=\omega}=\beta \omega\left(1-\frac{   \Omega^2_T}{  \omega^2 }\right) \ , \label{alpha-1} 
\end{equation}
\begin{equation}
\gamma\equiv {\it \Gamma}_{\epsilon_j=\omega}= q f(\beta\omega)\frac{\beta   \Omega^4_T}{N \omega^3  }.
\end{equation} 
We introduced here  the collective Rabi frequency renormalized by $T$, 
\begin{equation}
\Omega_T\equiv\Omega \sqrt{\tanh\frac{\omega }{2T}}  \ , \label{omega-T}
\end{equation}
and the function $f(x) = \frac{\sinh x - x}{1+ \cosh x  }\coth^2 \frac{x}{2} 
$; the parameter $q$ is a  ratio between fourth and second moments for coupling parameters, 
 $
    q = \langle g^4\rangle_j/\langle g^2\rangle^2_j 
    $.
The absence of the disorder in $g_j$ corresponds to $q=1$; in disordered case $q>1$; $q=9/5$ for a flat distribution ranging from $g_{\rm min}$ to $g_{\rm max}$  with  $g_{\rm max}\gg g_{\rm min}$.

In further consideration the  photon number \begin{equation}  
 \langle N_{\rm ph}\rangle=\beta^{-1}\int\limits_0^\beta \langle\bar\psi_\tau\psi_{\tau}\rangle d\tau \label{N-ph-def}
 \end{equation} is analyzed. Alternatively, it is given by the following identity 
\begin{equation}	\langle N_{\rm ph}\rangle=T\sum_n  (-G_n)-\frac{1}{2},  \label{Nph-1}
	\end{equation}
\begin{equation}	 G_n=-\beta\langle \bar \psi_n  \psi_n\rangle  \label{G-n-def}
	\end{equation}	
	 where $ G_n$ is $n$-th component of Matsubara Green function. If the quadratic expansion in (\ref{tr_log})  is applied then one obtains $S_{\rm eff, 0}[\Phi,\bar\psi_n,\psi_n]$ with $\gamma=0$. This  action is fully Gaussian with respect to all Matsubara modes. The following expression for the Green function is obtained for arbitrary $\epsilon_j$ and $\omega$ within this expansion:
\begin{equation}
	G_{n }=\frac{1}{{\rm i}2\pi n T-\omega-\Sigma_n[0]}  \ . \label{Gn-0}
\end{equation}
 For  the resonant limit  $\epsilon_j =\omega$ it reads:
\begin{equation}
	G_{n }(\epsilon_j {=}\omega)=\frac{\omega-2{\rm i}\pi n T}{(2 \pi n T+{\rm i}\omega)^2+ \Omega^2_T} \	. \label{Gn}
\end{equation}
It is used in the calculations below. 
This expression holds for   any $n$ in the Gaussian approach ($\gamma=0$).  After the summation one obtains the average photon number:
\begin{multline}
\langle N_{\rm ph } \rangle_{\rm   Gauss} =\\ =\frac{1}{4}\left[\coth\frac{\omega- \Omega_T}{2T}+\coth\frac{\omega+ \Omega_T}{2T} \right]-\frac{1}{2} \ . \label{NphGauss}
\end{multline}

 One can see that $\langle N_{\rm ph } \rangle_{\rm   Gauss}$ is divergent at the critical value of the renormalized Rabi frequency $ \Omega_{T,c}=\omega$ and is negative for $ \Omega_T>\omega$.	 This  follows from the condition  $G_{n=0}=-\frac{1}{\alpha T}$. It is divergent at the critical point where $\alpha= 0$.
 The regularization is provided by the expansion with respect to $\mathcal{G}$ in (\ref{tr_log}) which involves high order terms by $\Phi$. As we have shown above, quadratic expansion of $S_{\rm eff}$ by $\Phi^2$ gives $S_{\rm eff,0}$.    Corresponding zero mode's Green function is changed to
	$ 
	G_0=-\frac{\langle\Phi\rangle}{T}
	$ which is not divergent anymore at the critical point due to   $\gamma\neq 0$.  
	
	Within the $S_{\rm eff,0}$ action, for non-zero modes the expressions for $ G_{n\neq 0}$ are  the same as in (\ref{Gn}). 
	 We refine the  definition for the average $\langle N_{\rm ph } \rangle$ as  
	\begin{equation}	\langle N_{\rm ph } \rangle =\langle\Phi\rangle+\sum\limits_{n\neq 0} \langle \bar \psi_n  \psi_n\rangle-\frac{1}{2}. \label{Nph-2}
	\end{equation}
	The zero mode' part is written explicitly here. We emphasize thereby that it is calculated within the non-Gaussian (fourth order) approach by $\psi_0$.
Let us calculate both of the contributions originating from the superradiant order parameter, $\langle\Phi\rangle$, and from the thermal excitations $\langle \bar \psi_n  \psi_n\rangle$.  As long as there is no explicit dependence on $\varphi$, one has $\iint d (\re \psi_0) \, d (\im \psi_0) = \pi \int\limits_0^\infty d\Phi$. For $\langle\Phi\rangle$ we find
	\begin{equation}
		\langle\Phi\rangle 
		=\frac{\int\limits_0^\infty \Phi e^{ -S_{\rm zm}[\Phi ]} d\Phi}{\int\limits_0^\infty   e^{ -S_{\rm zm}[\Phi ]} d\Phi}=-\frac{\alpha}{2\gamma}+\frac{e^{-\frac{\alpha^2}{4\gamma}}}{\sqrt{\pi \gamma} { \rm erfc}\frac{\alpha}{2\sqrt{\gamma}}} \label{Phi}
	\end{equation}
	with the complementary error function is ${\rm erfc}z=1-{\rm erf}z$.
Summation over $n\neq 0$ gives the quasiparticle contribution  
\begin{equation}
	\sum\limits_{n\neq 0} \langle \bar \psi_n  \psi_n\rangle=\langle N_{\rm ph } \rangle_{\rm   Gauss}-\frac{1}{\alpha}.	\label{n:fluct}
\end{equation}
Finally, for the average photon number (\ref{Nph-2}) we obtain 
\begin{equation}
\langle N_{\rm ph } \rangle=-\frac{\alpha}{2\gamma}+\frac{e^{-\frac{\alpha^2}{4\gamma}}}{\sqrt{\pi \gamma} { \rm erfc}\frac{\alpha}{2\sqrt{\gamma}}}+\langle N_{\rm ph } \rangle_{\rm   Gauss}-\frac{1}{\alpha}. \label{n-ph}
\end{equation}

The fluctuations of the photon number are given by the second cumulant $\langle\!\langle N_{\rm ph}^2\rangle\!\rangle\equiv \langle N_{\rm ph}^2\rangle-\langle N_{\rm ph}\rangle^2$. With use of the above notations it  is reduced to
\begin{equation}
	\langle\!\langle N_{\rm ph}^2\rangle\!\rangle= \langle  \Phi^2\rangle -\langle  \Phi\rangle^2 +T^2\sum\limits_{n\neq 0} G_n^2.
\end{equation}
Calculation of the integrals by $\Phi$  and summation over $n$ provide
\begin{multline}
	\langle\!\langle N_{\rm ph}^2\rangle\!\rangle=  \frac{1}{2\gamma}+\frac{  \frac{\sqrt\pi \alpha}{2\sqrt\gamma } { \  \rm erfc} \frac{\alpha}{2\sqrt{\gamma}}-e^{- \frac{\alpha^2}{4\gamma}} }{  \pi \gamma  { \  \rm erfc}^2\frac{\alpha}{2\sqrt{\gamma}}}e^{-\frac{\alpha^2}{4\gamma}}+\\+\langle\!\langle N_{\rm ph }^2 \rangle\!\rangle_{\rm   Gauss}-\frac{1}{\alpha^2} \ .  \label{c2-ph}
\end{multline}
We introduced here the second cumulant in Gaussian approximation $\langle\!\langle N_{\rm ph}^2\rangle\!\rangle=T^2\sum\limits_{n} G_n^2$. It reads
\begin{multline}
	\langle\!\langle N_{\rm ph}^2\rangle\!\rangle_{\rm Gauss}=\\ \frac{ \cosh\frac{\omega}{T}\left(\cosh\frac{\Omega_T}{T}+\frac{T}{\Omega_T}\sinh\frac{\Omega_T}{T}\right)-1-\frac{T}{2\Omega_T}\sinh\frac{2\Omega_T}{T}}{4\left(\cosh\frac{\omega}{T}-\cosh\frac{\Omega_T}{T}\right)^2}.
\end{multline}
Similar to (\ref{n-ph}), the divergent zero frequency  term in the sum is canceled by $1/\alpha^2$ in (\ref{c2-ph}).

In the  Section \ref{sec:ph_tr} the  properties of the photon number and its second cumulant near the phase transition are analyzed in   details.

\section{ Phase  transition at low   temperatures. Resonant limit}\label{sec:ph_tr}
\subsection{Average photon number near the phase transition}

   In the following consideration   at low temperatures $T\ll \omega$, we should emphasize that  there is also a limitation  (\ref{condition-0}) which means that $T$ can not be arbitrary small. Namely, it  belongs to the domain
\begin{equation}
\omega\gg	T \gg \frac{\omega}{N} \ .  \label{condition-1}
\end{equation}
In such a limit we set $f(\beta\omega)=1$ and $\Omega_T=\Omega$ with the exponential by $\beta\omega$ accuracy. In this Section we continue to study the case of full resonance between cavity and all the qubits.

We obtain an analytical expansion of   photon number $\langle N_{\rm ph}\rangle$ (\ref{n-ph}) around the critical point $\Omega_{\rm c}=\omega$.  The expansion in  series by the dimensionless detuning  $\frac{   \Omega-\omega }{\omega}$ for $\langle N_{\rm ph}\rangle$ is 
\begin{multline}
		\langle N_{\rm ph}\rangle\approx\left[ \sqrt{\frac{NT}{\pi q\omega}}+\delta n_0\right]-\frac{1}{2} + \\+ \left[\frac{N(\pi - 2)}{\pi q }+\delta n_1\right]\frac{   \Omega-\omega }{\omega}+O\left[\left(\frac{   \Omega-\omega }{\omega}\right)^2\right] \ .
		 \label{n-ph-expansion}
\end{multline}

 The main contribution to   $\langle N_{\rm ph}\rangle$  follows from the $\psi_0$  mode   as powers of $\sqrt{\frac{NT}{\omega}}$.
The prefactors contain the leading term  given by zero mode, and small corrections $\delta n$. The corrections follow from the fluctuations of the   modes $ \psi_{n\neq 0}$.  Their expressions might be obtained  from the expansion of   (\ref{n:fluct}) as 
\begin{equation}
	\delta n_0=\frac{1}{4}-\frac{T}{4\omega}  \label{delta-n-0}
	\end{equation}
and 
\begin{equation}
	\delta n_1=\frac{3T^2-\omega^2}{24 T\omega} \ . \label{delta-n-1}
\end{equation}
The zero order term in (\ref{n-ph-expansion}) gives large but finite  photon number at the critical point  \begin{equation}\langle N_{{\rm ph} }\rangle_{\rm c}=\sqrt{\frac{NT}{\pi q\omega}}-\frac{1}{4}. \label{Nc}
\end{equation}

 The leading term is much higher than unity under the condition  $N\gg \omega/T$. If one goes beyond the validity of $S_{\rm eff,0}$   taking  a formal limit of $T\to 0$ in (\ref{Nc}) then   the  unphysical value of  $-\sfrac{1}{4}$
   is obtained.  This  demonstrates   that for low temperatures the  non-Gaussian fluctuations  are needed to  be taken into account. 
   It is known that $\langle N_{\rm ph} \rangle=1/2$ at zero temperature limit   above the critical point. 
  This is due to  the ground state wave function contains  \sfrac{1}{2}
    photon  on the  average. The ground state is changed at the critical point from a direct product of zero photon state and qubits ground state, $|n{=}0; \sigma_j{=}-1 \rangle$ ($j=1, \ ... \ N$), to an entangled state with a single  photon and excited qubits. 
  The field-theoretical approach provided does not allow to describe this limit because it is restricted to finite temperatures $T\gg \omega/N$. Nevertheless, this formalism allows to demonstrate a positive change  in  the negative constant value in  $\langle N_{{\rm ph} }\rangle_{\rm c}$ for very low $T$. For instance, if the third order correction $ \propto  \Phi \bar\psi_n \psi_n$  is taken into account in the action   $ S_{\rm eff,0} $.  
This correction originates from the dependence of the normal part of the self-energy $\Sigma_n $  on $\Phi $. In Appendix \ref{app-corr} it is shown that the integration over all non-zero modes provides a correction $\delta S[\Phi]= \delta\alpha   \Phi$ to $S_{\rm zm}[\Phi]$ due to the third-order term. At the critical point and  low temperatures $T\ll \omega$  this coefficient reads as  $\delta\alpha_{\rm c}=\frac{3 \omega}{4T N}$.  This results in the positive shift  in (\ref{Nc}) as   $\langle N_{{\rm ph} }\rangle_{\rm c}'=\langle N_{{\rm ph} }\rangle_{\rm c}+b$ where  $b=\frac{3}{8}(1-2/\pi)\approx 0.1363$.

\subsection{Fluctuations and the Fano factor  
}
  At the critical point and low temperature limit (\ref{condition-1}) the fluctuations of photons number are:
\begin{equation}  
	\langle\!\langle 
	N_{\rm ph}^2\rangle\!\rangle_{\rm c}= \frac{(\pi -2) N T}{2 \pi  \omega } + O[T/\omega]\ . \label{N2c}
\end{equation} 
This is the sum of large leading term due to superradiant order parameter fluctuations,  $\langle\!\langle \Phi^2 \rangle\!\rangle$, and  small correction $\sim T/\omega\ll 1$ due to weak fluctuations of quasiparticles.
The relative value of fluctuations
\begin{equation}
	 r=\frac{ \langle\!\langle 
	 	N_{\rm ph}^2\rangle\!\rangle  }{\langle N_{\rm ph}\rangle^2}, \label{r}
\end{equation}
is large as $e^{\omega/T}$ in the decoupling limit $\Omega=0$ and decays monotonously due to the decreasing of the second cumulant. It  is  less than unity above  the phase transition. Using the expressions for $\langle 
N_{\rm ph}\rangle$ and $\langle\!\langle 
N_{\rm ph}^2\rangle\!\rangle$, Eqs. (\ref{n-ph}) and (\ref{c2-ph}),   one obtains the expansion near the phase transition  up to the first order by the dimensionless detuning:
   \begin{equation}
	r\approx \frac{\pi-2}{2}
		 -  (\pi-3) \sqrt{\frac{\pi N \omega }{ qT}} \ \frac{   \Omega-\omega }{\omega}  \ .  
	\label{r-1}
\end{equation}  At the critical point ($\Omega=\omega$) the main contribution is due to the zero mode and, consequently, $r_{\rm c}=\frac{\langle\!\langle \Phi^2 \rangle\!\rangle}{\langle \Phi\rangle^2}$. 
     The universal value of the relative fluctuations $r$ at the transition point is \begin{equation}r_{\rm c}=\frac{\pi}{2}-1.\end{equation}
   It follows from the $\Phi$-integrals (\ref{Phi}) at $\alpha=0$  and is exact up  to the small correction $\sim N^{-1/2
   }$.

    From the expansion  (\ref{r-1}) 
    the width   of the fluctuational Ginzburg-Levanyuk region, $\Omega_{\rm GL}$, near  the critical  Rabi frequency  can be defined. This is a domain where fluctuations and average value  of the number of photons  are of the same order. This consideration can be applied straightforwardly to  the superradiant phase where $\Omega>\Omega_{\rm c}$. 
     The parameter $\Omega_{\rm GL}$ is obtained from the matching conditions
    \begin{equation}
    	r(\Omega)\sim 1 , \  \Omega-\Omega_{\rm c}\sim \Omega_{\rm GL} \ ,
  \label{fluct-def}
    \end{equation} 
    which give 
    \begin{equation}
    	\Omega_{\rm GL}\sim\sqrt{\frac{\omega T}{N}}. \label{fluct-zone}
    \end{equation}
  
 Approaching the critical point  from the normal phase, i.e.,  $\Omega<\Omega_{\rm c}$,  fluctuations are always greater that   average values and the definition (\ref{fluct-def}) is not valid. Instead of (\ref{fluct-def}) we introduce the width $\Omega'$  where the  superradiant order parameter fluctuations start to grow and become relevant. In this region the  contribution to $\langle N_{\rm ph }\rangle$ due to the non-Gaussian fluctuations of  $|\psi_0|^4$ is comparable with the quasiparticle's part related to $\psi_{n \neq 0}$.
 We define $\Omega'$ through the value of $\Omega=\Omega_{\rm c}-\Omega'$ which provides the matching  between the average values obtained in the Gaussian and non-Gaussian approaches:
  \begin{equation}
 	\langle N_{\rm ph }\rangle_{\rm   Gauss}\sim \langle N_{\rm ph }\rangle ,  \  \Omega_{\rm c}- \Omega\sim \Omega'  \ . \label{fluct-def-normal}
 \end{equation}
From   (\ref{NphGauss}) and (\ref{n-ph}) it follows that 
 $ \langle N_{\rm ph }\rangle_{\rm   Gauss}\sim  T/\Omega'$ and $\langle N_{\rm ph }\rangle\sim \sqrt{NT/ \omega}$. The width $\Omega'$  of fluctuation-dominated region appears the same order as in the superradiant phase, i.e.,
 \begin{equation}
 	\Omega'\sim\Omega_{\rm GL}\sim\sqrt{\frac{\omega T}{N}} \ . \label{fluct-zone-normal}
 \end{equation}
   It is rather narrow and is much less than the temperature due to  the condition (\ref{condition-1}).

 \begin{figure*}[htp]
    \includegraphics[scale=0.48]{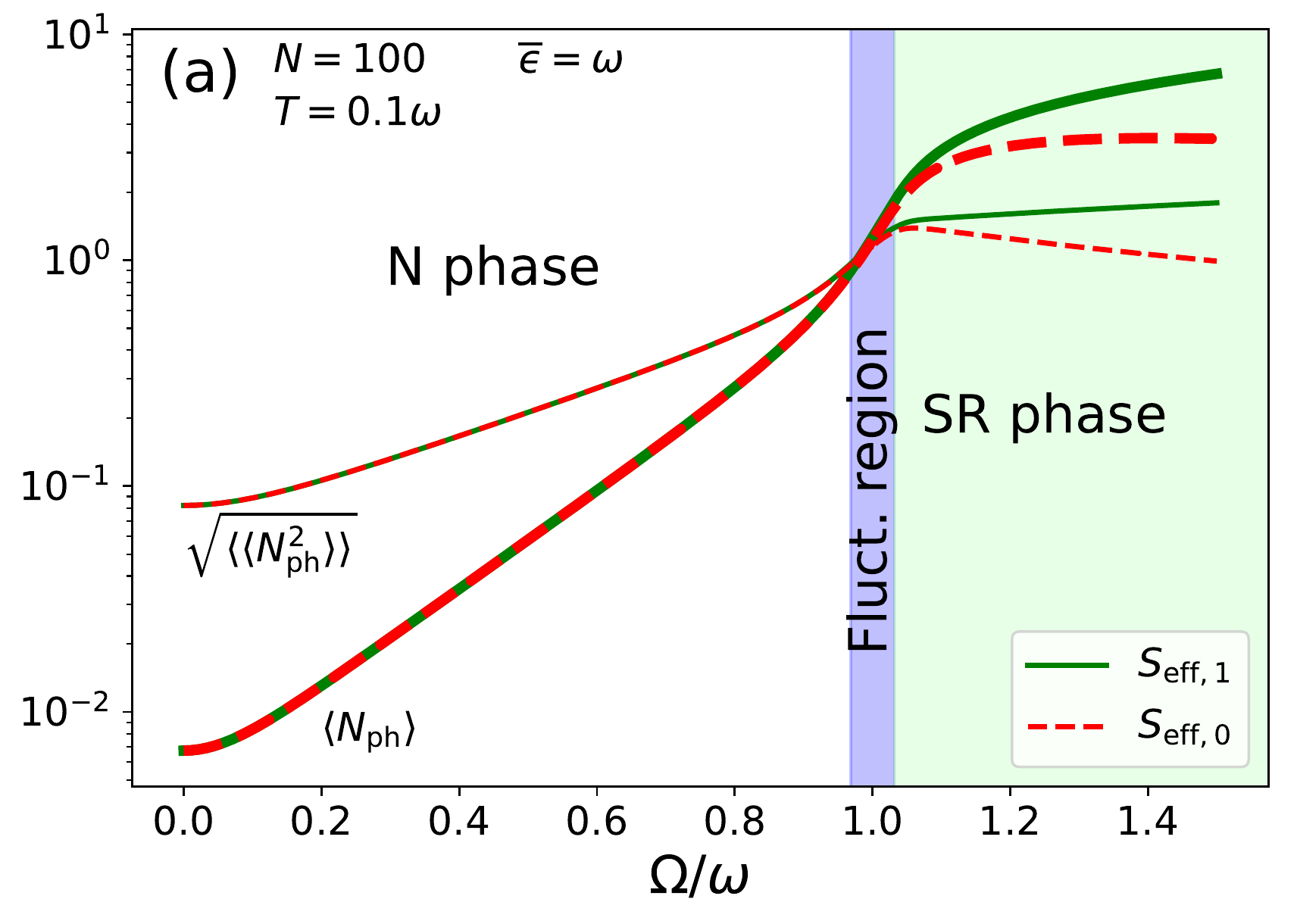}
    \includegraphics[scale=0.48]{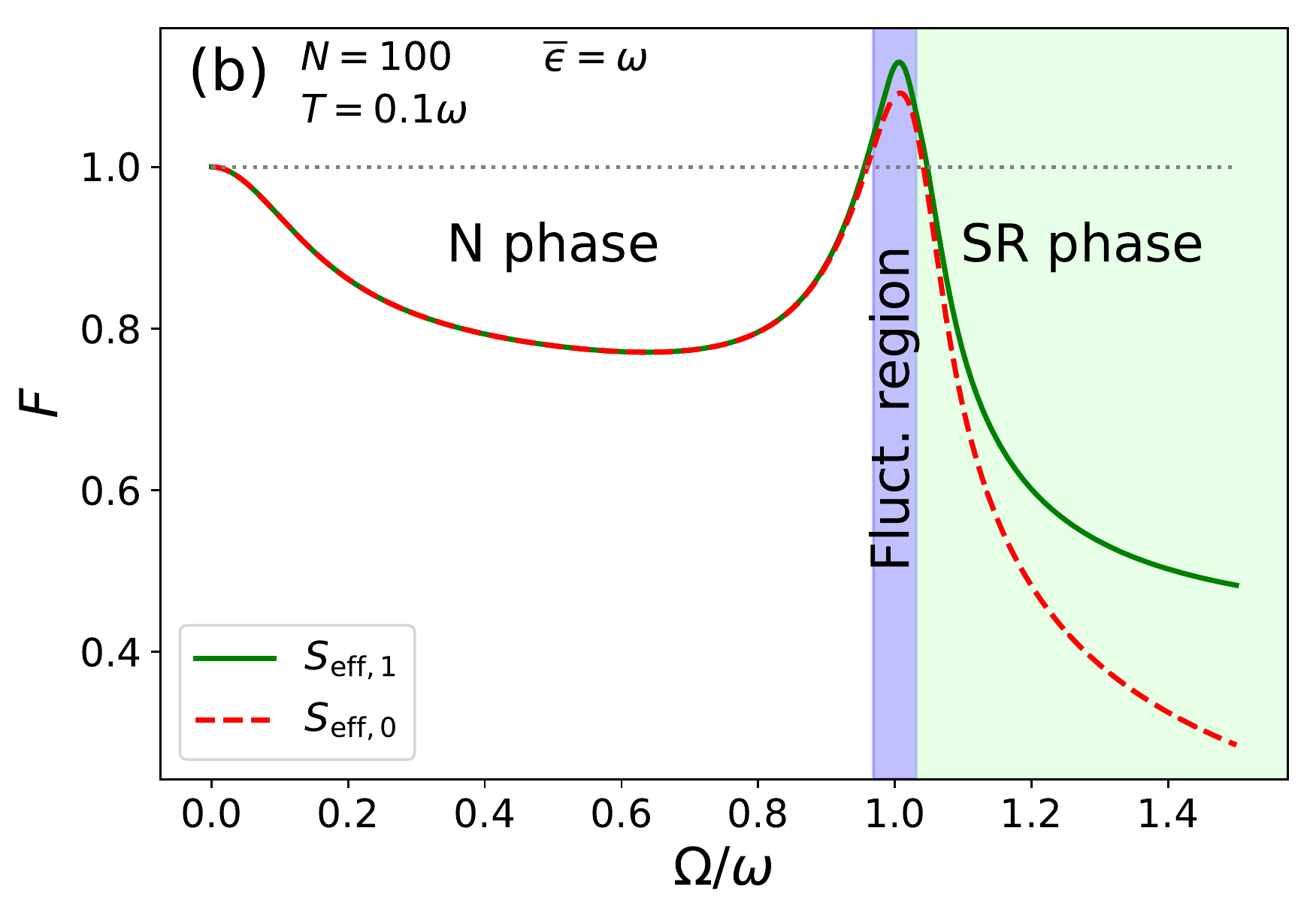}
    \\
    \includegraphics[scale=0.48]{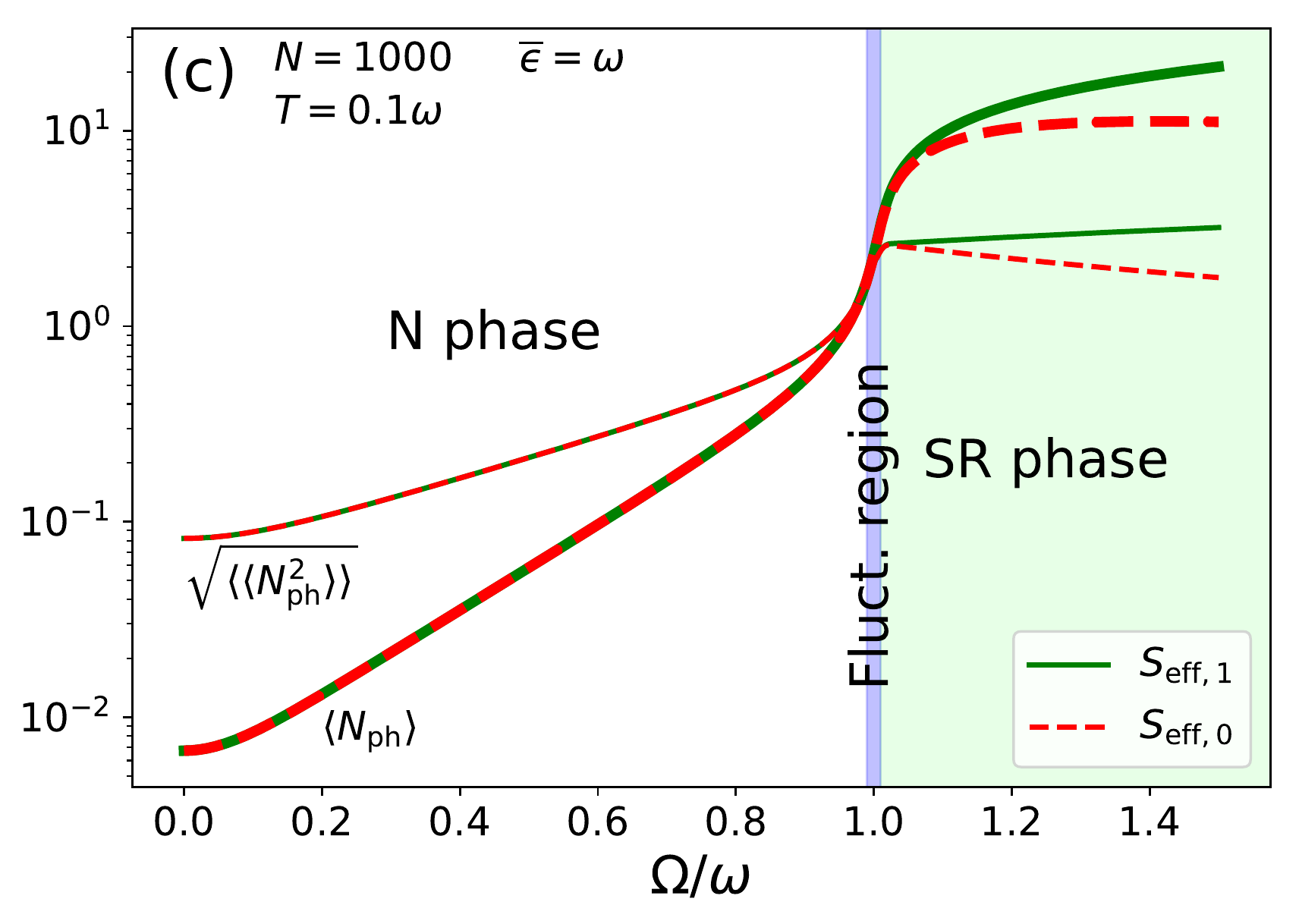}
    \includegraphics[scale=0.48]{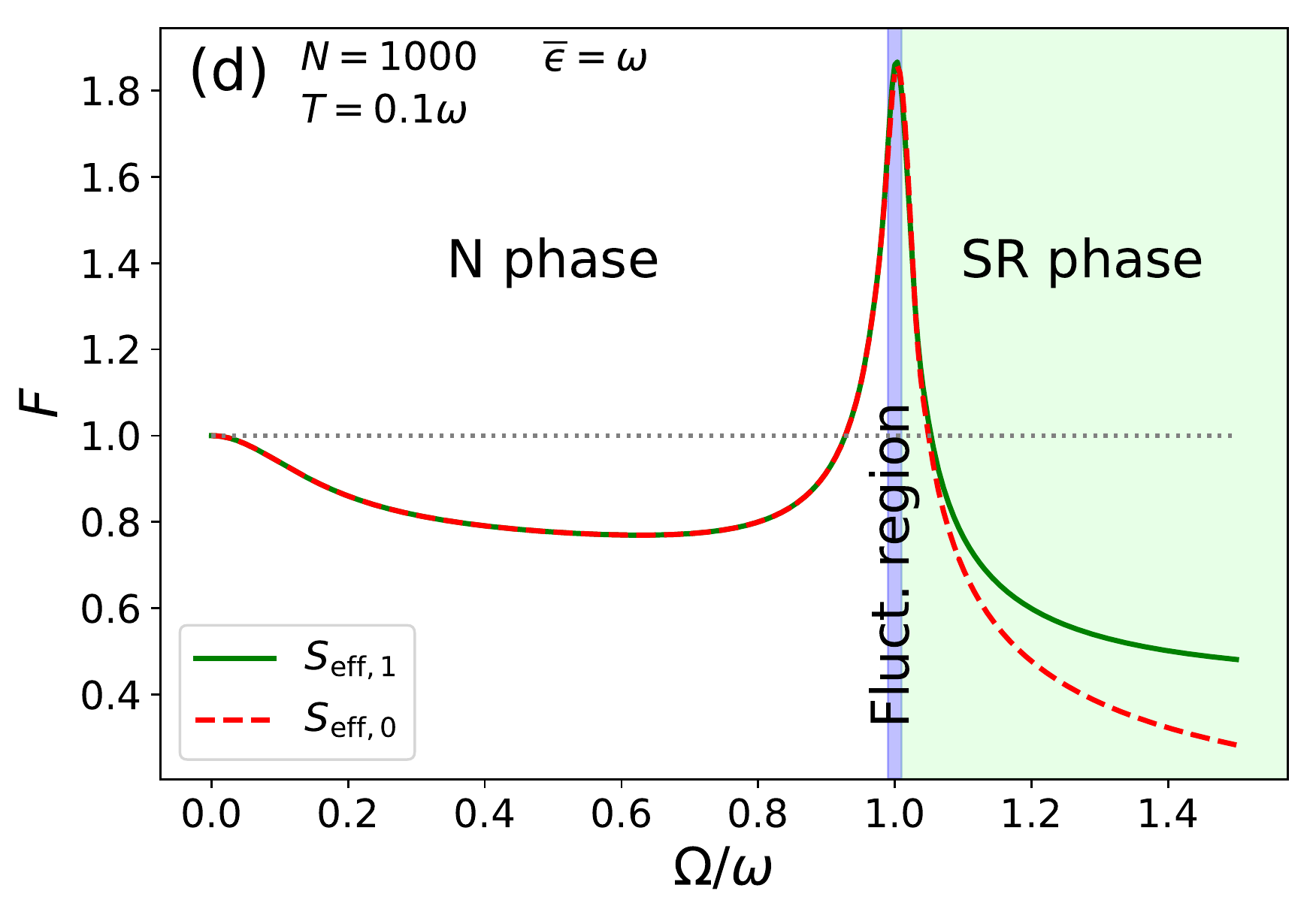}
    \caption{  
            (a, c) Average photon number $\langle N_{\rm ph}\rangle$, fluctuations $\sqrt{\langle\!\langle N_{\rm ph}^2\rangle\!\rangle}$ and (b, d) Fano factor  $F$ as functions of collective Rabi frequency $\Omega= g\sqrt{N }$ (in  units of resonator mode frequency $\omega$).  All curves  are calculated for the full  resonance limit between  qubits and cavity mode   frequencies, $\bar\epsilon=\omega$. The temperature is low, $T=0.1 \, \omega$, and qubit number is $N=100$ on   panels (a, b) and $N=1000$ on panels (c, d). White, light blue and light green areas (color online) correspond to normal (N) phase, fluctuational region and superradiant (SR) phase, respectively. The critical point  is $\Omega_{\rm c}=  \omega $  and the width of the fluctuational region is $2\Omega_{\rm GL}\approx 0.063 \ \omega$ on panels (a, b) and $2\Omega_{\rm GL}= 0.02 \ \omega$ on panels (c, d). Red and green curves stand  for calculations based on $S_{\rm eff,0}$ and $S_{\rm eff,1}$ actions, respectively. The  Fano factor   (b, d) in the normal phase demonstrates that  $F_{\rm min}<F<1$.  It means negative correlation between photons (antibunching effect). The horizontal dotted line $F=1$  separates the regions of negative ($F<1$) and positive ($F>1$) photon correlations. The fluctuational region in (b, d) demonstrates a growth of the Fano factor with  a peak at $F_{\rm c}>1$ which means positive correlations between photons. The superradiant phase shows reentrance to the negative correlations    with the decay of $F$.}   \label{plots}
\end{figure*}

In order to illustrate the above results we present  in  Fig.~\ref{plots} (a, c)    the data for   $\langle N_{\rm ph}\rangle$ and    $\sqrt{\langle\!\langle N_{\rm ph}^2\rangle\!\rangle}$   as functions of $\Omega$. We consider the full  resonance limit, $\bar\epsilon=\omega$, and low temperature regime   $T=0.1 \ \omega$. The qubits number is $N=100$ on panel (a) and $N=1000$ on panel (c), hence, the constraint (\ref{condition-1}) is satisfied. Such a qubit number can be realized in contemporary quantum metamaterials. White, thin light blue and light green sectors  correspond to normal (N) phase, fluctuational region and superradiant (SR) phase, respectively. The critical point in this low temperature regime  is $\Omega_{\rm c}=   \omega $. The width of the Ginzburg-Levanyuk fluctuational region is $2\Omega_{\rm GL}\approx 0.063 \ \omega$.  The    red curves are obtained with the use of the action $S_{\rm eff,0}$ and the corresponding analytical   results (\ref{n-ph}) and (\ref{c2-ph}). 
   It is shown that in the normal phase there are exponential dependencies of the photon number and its fluctuations, as follows from linear sectors in the logarithmic scale. Tuning $\Omega$ to the critical value  initiates  the superradiant transition where photons number is increased rapidly. Further increase of $\Omega$  drives the system into superradiant state. The  quadratic expansion for the logarithm in $S_{\mathcal{G}}$, as it should be, does not work well  in this phase. A  correct description assumes that the use of the   action $S_{\rm eff,1}$ from (\ref{s_eff1}).
    The green curves are the results obtained  by means of  $S_{\rm eff,1}$  where the integration  over  $\Phi$ is performed numerically.    Dashed parts of the red curves  demonstrate  the  difference between these two approaches. Green curves show that  $\langle N_{\rm ph}\rangle$ and its fluctuations   in the superradiant phase grow sub-exponentially. It also follows from this plot that  in the normal phase  the relative fluctuations value 
  $r_{\rm n}>r_{\rm c}$ and $r_{\rm sr}<r_{\rm c}$ in the superradiant phase.

   It is also instructive to analyze the Fano factor 
   defined as  a  ratio between second and first cumulants as
   $$
   F=\frac{\langle\!\langle 
   	N_{\rm ph}^2\rangle\!\rangle}{\langle N_{\rm ph}\rangle}.
   $$ 
This is  a representative parameter   bringing an information about the statistics. 
   The value of  $F$ reflects a type of a coherence between the photons: $F=1$ means that they are uncorrelated, $F<1$ and $F>1$ correspond to their negative and positive correlations, respectively.  As shown in Fig.~\ref{plots} (b, d) the dependence   $F(\Omega)$ demonstrates rich behavior. The parameters of  calculation here are the same as that in     the plot (a):  $T=0.1 \ \omega$,  $N=100$ and $\bar\epsilon=\omega$. Red and green curves correspond to calculations based on   $S_{\rm eff,0}$ and $S_{\rm eff,1}$, respectively.
   In the  decoupling limit $\Omega=0$  the    value of the Fano factor is
  \begin{equation}
   	F_0=\frac{1}{1-e^{-\beta \omega } } >1  \ . \label{F-0}
   \end{equation}
  In a low temperature limit  $F_0\approx 1+e^{-\beta \omega }$ which means that photons are weakly correlated. 
      For a finite $\Omega$ there is the entrance   into the negative correlations domain where the  dependence is non-monotonous with  $F_{\rm min}<F<1$.  There is a minimum with $F_{\rm min}\approx 0.8$ for an intermediate strength of $\Omega$. It means negative correlation between photons (antibunching effect) in the normal phase due to the interaction  between photons  through the qubit environment. 
      It is remarkable, that the dependence $F(\Omega)$ in the fluctuational region    demonstrates a dramatic  change where  $F$ grows rapidly and becomes greater than unity. There is a maximum at the critical point which is given by
    \begin{equation}
    	F_{\rm c}=\frac{1}{2}(\pi-2) \sqrt{\frac{NT}{\pi \omega}}  \ .
    \end{equation} The latter means strongly positive coherence between photons near the superradiant transition. The Fano factor shows  the decay entering into the superradiant phase if $\Omega$ is further increased. As one can see there is the reentrance to negative correlations with $F_{\rm sr}<1$. 
    
    The finite width of  fluctuational region and the peak in the Fano factor dependence are  finite-size effects. In thermodynamic limit of $N\to \infty$ the Fano factor peak shrinks to a  singularity at the critical point. This tendency is seen from a comparison of Figs.~\ref{plots} (b) and  (d) where $N$ is changed by an order.

\subsection{Numerical simulation}
\begin{figure*}[htp]
	\includegraphics[scale=0.45]{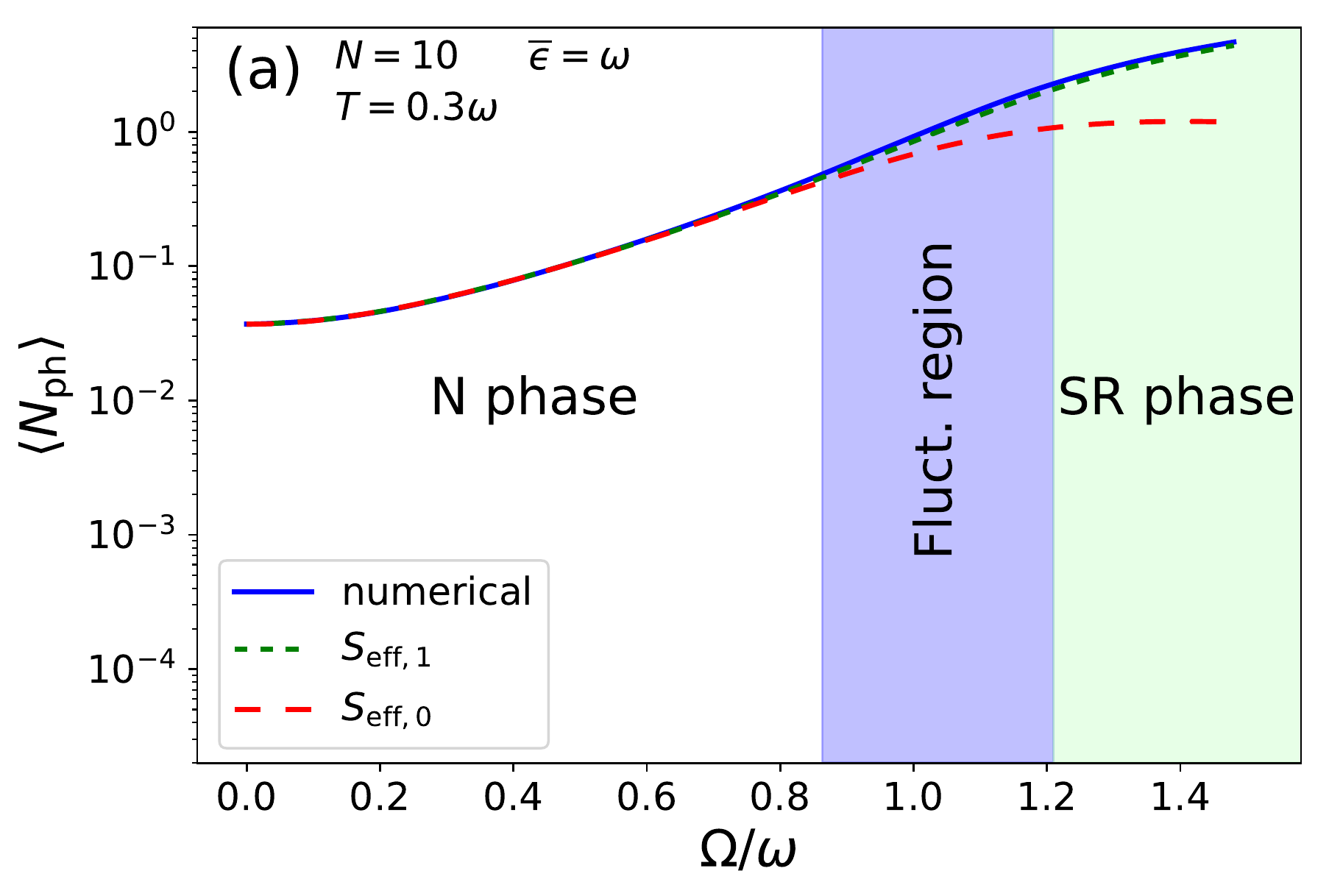}
	\includegraphics[scale=0.45]{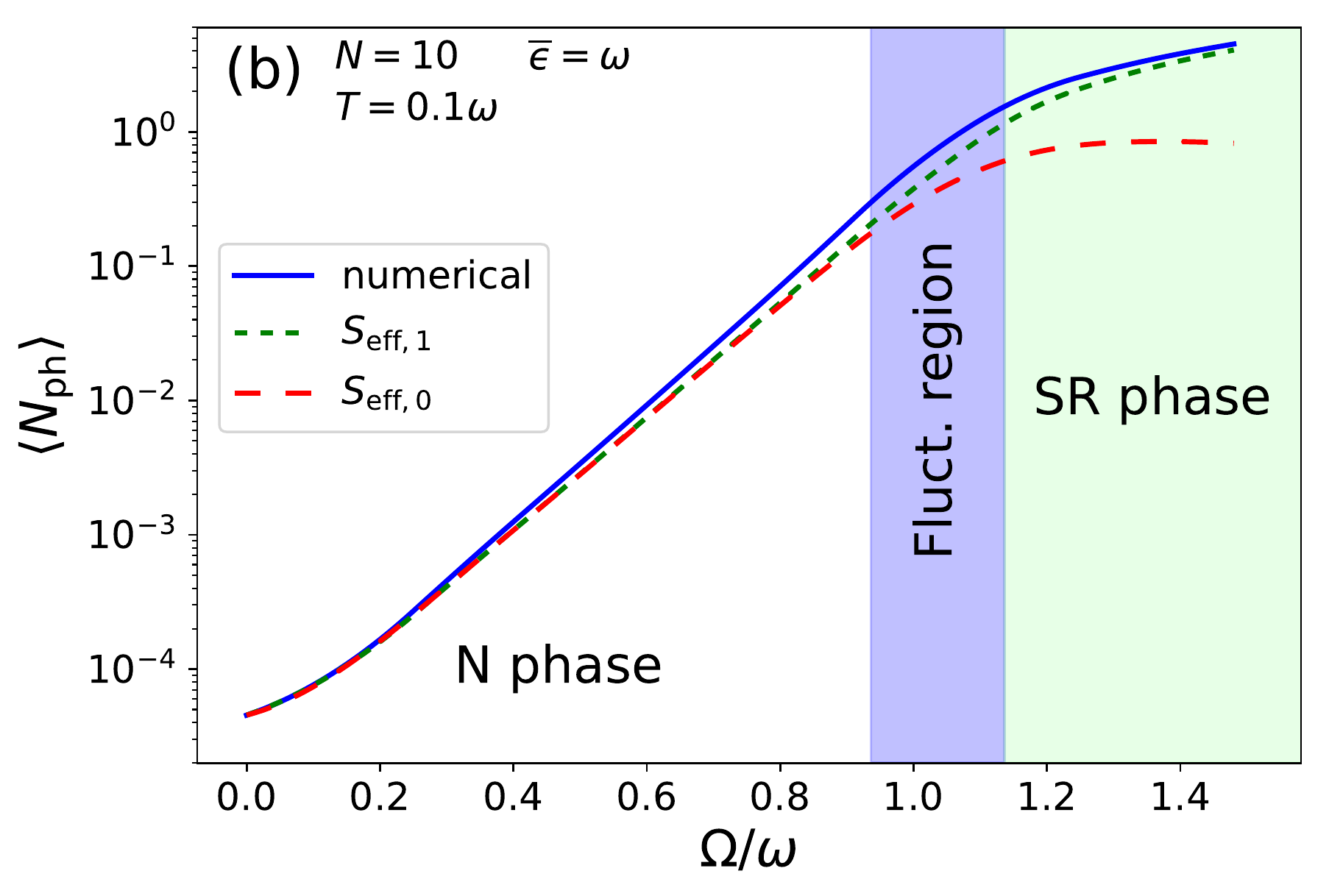}
	\caption{
		   Comparison of results obtained in effective action techniques and in exact numerical calculations based on equilibrium density matrix. The data for the average photon number $\langle N_{\rm ph}\rangle$    
		is presented.  
		The temperature is low as  $T/\omega=0.3$  for the panel (a) and $T/\omega=0.1$ for the panel (b) and qubit number   is $N = 10$.  The range of cubit-cavity coupling covers the domain of the normal phase, fluctuational region and the   superradiant phase. The data obtained from numerical simulations is shown as blue curves. Results of field theoretical approaches based on $S_{\rm eff,0}$ and $S_{\rm eff,1}$ are shown as red  {dashed} and green  {dotted} curves, respectively.    We note surprisingly small deviation of solid blue lines from the green dotted line. Despite that  the parameters are near the edge of applicability range of the theory, a good agreement between numerical results and theoretical calculations  is clearly observed. 
		 	} \label{plots-num}
\end{figure*}

In Fig.~\ref{plots-num} we compare the results obtained in   the field-theoretical formalism  and that in exact numerical simulations. The qubit number $N=10$  means that the system is   beyond from the thermodynamic limit.  
Despite that $\kappa_{\rm c}$ is not very small compared to unity,  $\kappa_{\rm c}\approx \ 0.58$ on Fig.~\ref{plots-num} (b), a well quantitative agreement  between the numerical and  theoretical results is observed.  Surprisingly, analytical solution is in a good agreement with numerical calculations  even when  $\kappa_{\rm c}=1$, as shown in Fig.~\ref{plots-num} (b). We represent results for $\langle N_{\rm ph} \rangle$ and $\langle\!\langle N_{\rm ph}^2\rangle\!\rangle$ obtained in three different ways.     The  red dotted  curves  are  obtained with the use of the action $S_{\rm eff,0}$ and  analytical  expressions (\ref{n-ph}) and (\ref{c2-ph}).
Green  dashed  curves are derived with the use of $S_{\rm eff,1}$. There is a difference between  them in the superradiant  phase 
Blue  solid  curves represent  the results of  numerical calculations based on the  definitions
\begin{equation}
	\langle N_{\rm ph} \rangle=\frac{{\rm Tr}[\hat\rho \hat\psi^{\dagger}\hat\psi]}{{\rm Tr}[\hat\rho ]} \ ,
\end{equation} 
	and
\begin{equation}	
	\langle\!\langle N_{\rm ph}^2 \rangle\!\rangle=\frac{{\rm Tr}[\hat\rho  (\hat\psi^{\dagger}\hat\psi)^2]}{{\rm Tr}[\hat\rho ]} -\langle N_{\rm ph} \rangle^2 \ .
\end{equation} 
Here the equilibrium  density matrix is 
\begin{equation}\hat\rho=\exp(-\hat H/T)\ . 
\end{equation}
It is block diagonal due to the conservation of total excitations number in the system. This follows from a commutation of the excitations number operator, $\hat M =\hat\psi^\dagger\hat\psi+\sum_j\hat\sigma^+_j\hat\sigma^-_j$,  and $\hat H$. 
In calculations the maximum of excitations number is $M_{\rm max}=50$.  This  means that  $\hat\rho$ has $M_{\rm max}$ blocks each of them has the  dimension of $2^M$, $ M=1, \ ... \ ,   M_{\rm max}$. For the above parameters   the most  relevant part of  the Fock space belongs to  $M$ which covers a region  from one to a value around 30.
We observe a good  correspondence  between theoretical  curves (red  dashed and green dotted) and numerical simulation  (blue solid curves)  for the range of Rabi frequencies $0<\Omega\lesssim   \omega$ which covers  the normal phase  and fluctuational region.
In the superradiant phase, where $ \Omega \gtrsim \omega $, the  numerical results are in good agreement with more precise calculations based on $S_{\rm eff,1}$.

 \section{Some generalizations}\label{sec:generalization}
 \subsection{High temperatures}
 Below we discuss    results  obtained at the critical point for the high temperature regime   $T\gg \omega$. Note that the phase transition at $\alpha=0$ (see Eq. (\ref{alpha})) is given by the increased collective coupling:
\begin{equation}
 \Omega_{\rm c}=\sqrt{T\omega}.
  \end{equation}
We use   (\ref{n-ph}) and (\ref{c2-ph}) to obtain the leading order expansions for $\langle N_{\rm ph}\rangle$ and $\langle\!\langle 
N_{\rm ph}^2\rangle\!\rangle$ by the large parameter $T/\omega $. 

In the  Appendix \ref{app-corr} we discuss that the Gaussian approximation for quasiparticle fluctuations is valid for any  $N$ and  the corrections due to cross terms $\propto \Phi\bar\psi_n\psi_n$ are always small. This  is distinct from   $N\gg \omega/T\gg 1$ in the low-temperature limit addressed above. 

We obtain that the photon number at the critical point is
  \begin{equation}
\langle N_{\rm ph}\rangle_{\rm c}=\sqrt{\frac{3 N}{\pi }} \frac{T}{\omega}.
  \end{equation}
 In contrast to the low temperature limit where it scales as  $\propto \sqrt{T}$,  in the high temperature regime under consideration it grows as $\propto T$.
The fluctuations of photons,
\begin{equation}
\langle\!\langle 
N_{\rm ph}^2\rangle\!\rangle_{\rm c}= \frac{3(\pi -2) N  T^2 }{2 \pi  \omega ^2}+\frac{T^{5/2}}{8 \sqrt{2} \omega ^{5/2}}, \label{fluct-high-t}
  \end{equation}
 in contrast  to (\ref{N2c}),  contain not only the contributions from  $\Phi$ (first term), but  also  from the non-zero modes $\psi_n$  as well (second  term).  
Thus, the high temperature limit is distinct   in that sense that there are two domains of $N$ where fluctuations have different contributions. 
The first domain for $N$ is related to the thermodynamical limit of very large qubit number. It is given by (\ref{fluct-high-t}) as
 \begin{equation}
	N\gg \sqrt{\frac{T}{\omega}},
\end{equation}
 when only superradiant zero mode is relevant. 
 The second one is  the intermediate region,  
  \begin{equation}
    \sqrt{\frac{T}{\omega}} \gtrsim N  \  , \label{condition-high-t}
  \end{equation}
 when contribution of fluctuations of the order parameter can be neglected compared to that of thermal fluctuations of quasiparticles.
 The relative value at the transition for this  intermediate domain,
  \begin{equation}
 r_{\rm c}=\frac{\pi-2}{2}  +\frac{\pi  \sqrt{T}}{24 \sqrt{2}N \sqrt{\omega }} \ , \label{r-c-high-t}
 \end{equation}
 shows a deviation  from the universal value $\pi/2-1$  due to the second term. Thus, $N\sim\sqrt{ T / \omega} $ defines a condition for the crossover between two types of fluctuational behavior. Namely,  $N\gg \sqrt{ T / \omega} $ corresponds to thermodynamic limit   where fluctuations of  superradiant order parameter provide the leading contribution to fluctuations of the photons number. In case of $  \sqrt{ T / \omega}\gtrsim N$ the contribution due to  thermal fluctuations of quasiparticles becomes dominant.

 \subsection{Inhomogeneous broadening}
  In the above results for the   resonant limit a spread of coupling energies $g_j$ yields the prefactor  $q^{-1}$ for the qubit number. The inhomogeneous broadening of qubit energies modifies the expressions in a more  significant way described below. 
  
  We also assume that qubit frequencies are distributed in a certain interval,  temperatures are low enough, $T\ll \epsilon_j$, and couplings are homogeneous, $g_j\equiv g$. We assume that the system is in the critical point,  $\alpha=0$, and photons number (\ref{n-ph}) is  contributed by the  zero mode only, i.e., $\langle N_{\rm ph}\rangle = \frac{1}{\sqrt{\pi\gamma}}$, and   quasiparticles contributions are  neglected. In the definition for $\gamma$  (\ref{gamma}) the sum over qubit index is replaced by the integral over energies, $\sum_j\to  N \int \rho( {\epsilon}) \, d  {\epsilon}$ with the density of states   $\rho( {\epsilon})$ is normalized to unity ($\epsilon$  is a qubit's energy). We discuss two cases which correspond to flat distributions with finite and very broad widths. 
  
  In the first case we consider the distribution with a median energy at  $\overline \epsilon$  and width $\Delta$, hence, the density of states is 
  \begin{equation}
  	\rho({\epsilon})=\frac{1}{\Delta}\theta(\Delta/2-|{\epsilon}-\overline{ \epsilon}|).
  	\end{equation}
The photon number is obtained as 
  \begin{equation}
\langle 
N_{\rm ph} \rangle=z(\Delta/{\overline \epsilon})\frac{\sqrt{NT{\overline \epsilon}}}{\sqrt \pi \omega} , \label{n-ph-offres}
\end{equation}
where dimensionless prefactor $z$ is 
  \begin{equation}
	z(x)= \left(\frac{1}{x}-\frac{x}{4} \right) \ln \frac{1+x/2}{1-x/2}.
\end{equation}
  In the homogeneous limit, $\Delta\to 0$, this prefactor is unity. Note, that the expression (\ref{n-ph-offres}) provides the photon number at the critical point for the off-resonant regime, where ${\overline \epsilon} \neq \omega$.
  
  In the second case of the  very broad  distribution,   qubits energies belong to the interval from  $  \epsilon_{\rm min}$  up to  large $\epsilon_{\rm c} \gg \epsilon_{\rm min}$ which is  spectrum cut-off. This case is considered as a thermodynamic limit  where the average level spacing can be introduced, $\delta\epsilon\equiv \epsilon_{\rm c}/N$.  Under the assumption 
   \begin{equation}
  T\ll \{{  \epsilon_{\rm min}},\omega \} \ll \epsilon_{\rm c}
  \end{equation}
we find that 
  \begin{equation}
 	\langle 
 	N_{\rm ph} \rangle= \sqrt\frac{2T}{\pi \delta\epsilon} \frac{  \epsilon_{\rm min}}{\omega}\ln \frac{\epsilon_{\rm c}}{  \epsilon_{\rm min}} . \label{n-ph-wide}
 \end{equation} 
In a physically relevant  situation the lower edge of the qubits spectrum  $  \epsilon_{\rm min}$  may be of the order of the resonator mode frequency, hence, their fraction is order of unity. The logarithm is also   not a very large number.
Interestingly,  in this case we obtain that the photon number is affected mainly  by the ratio between   the smallest energy scales -- the temperature and level spacing.

\subsection{Off-resonant regime}
In this subsection we generalize the result for photon number where $\omega$ and $\bar \epsilon=\epsilon_j$ are out of the resonance. We assume no disorder in $g_j$. The value of $\langle N_{\rm ph}\rangle$ is given by the same expression as in Eq. (\ref{n-ph})  but $\alpha$, $\gamma $  and the Gaussian part  are taken in more general form due to $\bar\epsilon \neq \omega$. The functional coefficients are
\begin{equation}
	\alpha^{(\bar\epsilon{\neq}\omega)} =\frac{\omega}{T}-     \frac{\Omega^2 }{T \bar\epsilon}\tanh\frac{ \bar\epsilon}{2T} \ , 
	\label{alpha-off-res}
\end{equation}
\begin{equation}
\gamma^{(\bar\epsilon{\neq}\omega)}= \frac{\Omega^4  }{NT\bar\epsilon^3} \ \frac{\sinh \frac{ \bar\epsilon}{ T} - \frac{ \bar\epsilon}{ T}}{ \cosh\frac{ \bar\epsilon}{ T} +1 }, \label{gamma-off-res}
\end{equation}
The Gaussian part is given by the sum (\ref{Nph-1})  with $G_n$ from (\ref{Gn-0}). In the off-resonant case it reads as 
\begin{multline}
\langle N_{\rm ph}\rangle_{\rm Gauss}^{(\bar\epsilon{\neq}\omega)}= \\ =\frac{(\omega -\bar\epsilon ) \sinh  \frac{E(\bar\epsilon,\Omega)}{2 T} -E(\bar\epsilon,\Omega)\sinh  \frac{\omega +\bar\epsilon }{2 T} }{2 E(\bar\epsilon,\Omega) \left(\cosh  \frac{E(\bar\epsilon,\Omega)}{2 T} -\cosh  \frac{\omega +\bar\epsilon }{2 T} \right)}-\frac{1}{2} \label{NphGauss-off-res}
\end{multline}
where
\begin{equation}
E(\bar\epsilon,\Omega) = \sqrt{	4  \Omega^2 \tanh\frac{\bar\epsilon}{2T} +(\bar\epsilon -\omega )^2}.
	\end{equation}
In the resonant limit of $\bar\epsilon=\omega$, addressed in the Sec. \ref{sec:n_ph}, the expression (\ref{NphGauss-off-res}) reproduces  (\ref{NphGauss}). 

\begin{figure*}[ht]
 	\includegraphics[scale=0.52]{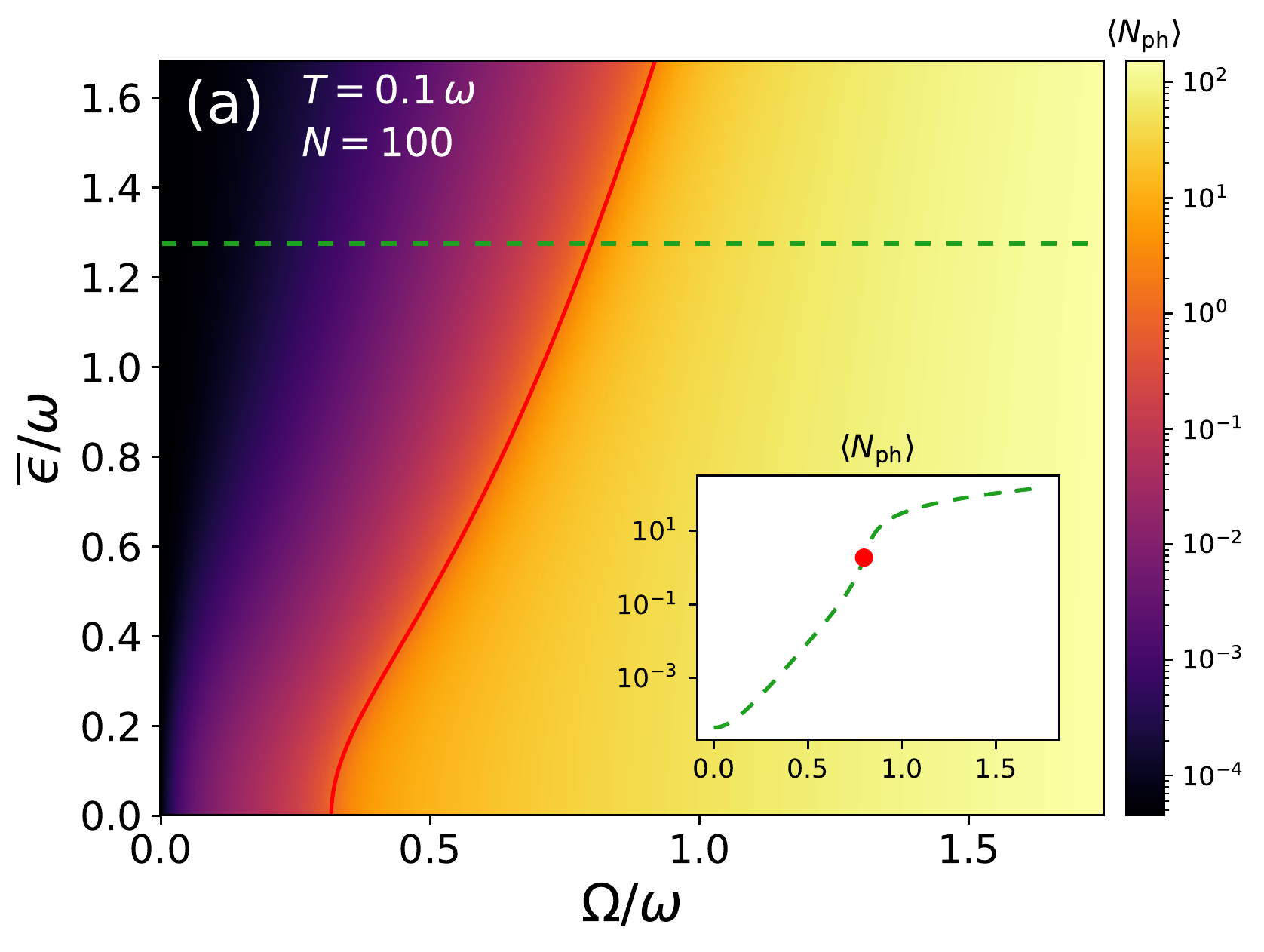}
 		\includegraphics[scale=0.52]{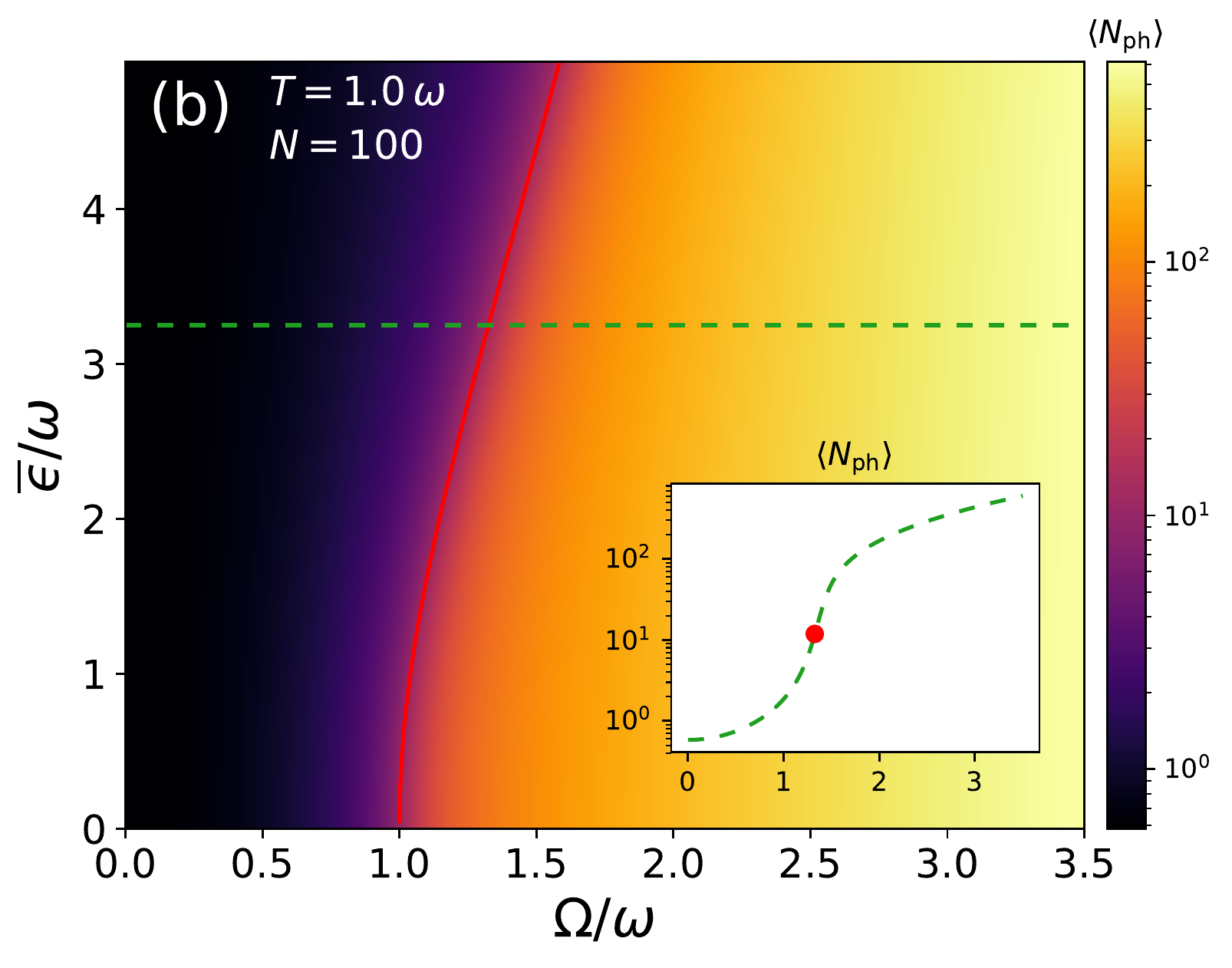}
 	\caption{  Average photon number $\langle N_{\rm ph}\rangle$ obtained by means of $S_{\rm eff,1}$   as the function of the Rabi frequency $\Omega$ and   qubit    energies  $\epsilon_j=\bar\epsilon$ in non-resonant regimes of $\omega\neq \bar\epsilon$.  The dark regions in the maps correspond to normal phase; bright regions correspond to superradiant phase. Qubit number $N=100$,   $\Omega$ and $\bar\epsilon$ are measured in   units of resonator frequency $\omega$. (a) Data calculated  for low temperature regime   $T=0.1 \, \omega$. (b) Data calculated  for intermediate temperature  $T= \omega$.   Red solid curve corresponds to the critical $\Omega_{\rm c}$ as a function of $\bar\epsilon$ given by the relation (\ref{omega-c}). Insets in (a) and (b) demonstrate $\langle N_{\rm ph}\rangle$  as functions of $\Omega/\omega$ for cuts   marked by green dashed lines; red points mark the critical  Rabi frequency for a given $T$ and $\bar\epsilon$ in the  cut. }
 	\label{maps}
 \end{figure*}

In Fig. \ref{maps}  (a) and (b) the photon number as the function of  $\Omega$ and   qubits   energies  $\bar\epsilon$  is plotted (in units of $\omega$). The effective action $S_{\rm eff,1}$ is employed in this calculation. The data shown in (a)  demonstrates the behavior at  low temperature $T=0.1 \, \omega$; (b)  demonstrates the behavior at  intermediate temperature $T=\omega$. The qubit  number $N=100$ in both of the plots.  The dark (bright) regions in the maps correspond to  normal (superradiant) phases. Red curves depict  dependencies of the critical coupling value $\Omega_{\rm c}(\bar\epsilon)$ from (\ref{omega-c}) where $\omega$ is kept constant.   Curves in   insets demonstrate  the average photon number as functions of $\Omega/\omega$ for cuts in the plots marked by green dashed lines. Red points in insets stand  for the critical  Rabi frequency for a given $T$ and cuts of $\bar\epsilon$ in (a) and (b). These plots demonstrate typical scales of photons number in normal and superradiant phases for low and intermediate temperature regimes. Red curves corresponding to $\Omega_{\rm c}(\bar\epsilon)$ relations reproduce asymptotics  for low temperatures in (a), where $\Omega_{\rm c}(\bar\epsilon)\propto \sqrt{\bar\epsilon}$, and for high temperatures with $\Omega_{\rm c}(\bar\epsilon)\propto {\rm const}$ in (b).
 
\section{ Full counting statistics}  \label{sec:fcs}

 \subsection{Generating action}
The effective action  for quantum fluctuations   (\ref{s_eff0}) allows to we derive the full counting statistics  (FCS)  for photon numbers.  These are cumulant and moment  generating functions (CGF and MGF). These are functions  of real counting variable $\xi$. In our consideration  the generating action is introduced  on the imaginary time.

The  CGF and MGF are defined as follows through the partition function $Z(\xi)$
 \begin{equation}
{\rm CGF}(\xi)= \ln {\rm MGF}(\xi) , \quad 	{\rm MGF}(\xi)=\frac{Z(\xi)}{Z(0)}, 
 \end{equation}
 \begin{multline}
 	Z(\xi)= \int D[\Psi] \exp\Big[-S_{\rm eff,0}[\Phi,\bar\psi_n,\psi_n]-\\-{\rm i}\xi\Big(\Phi+\sum\limits_{n\neq 0}  \bar \psi_n\psi_n-1/2\Big)\Big]. \label{z}
 \end{multline}
$\mathcal{T}$-ordering in the imaginary  time representation of the path integrals   
 assumes that the photon number, introduced in (\ref{N-ph-def}),  is defined as 
 \begin{equation}
 N_{\rm ph}=T\int\limits_0^\beta \bar\psi_\tau\psi_{\tau+{\it 0}} d\tau \label{n-def-0}
 \end{equation}
 in a generating term.  
Alternatively, the generating term can be also represented  as a half sum of (\ref{n-def-0}) with $+ {\it 0}$ and $- {\it 0}$, which is  symmetric under $\mathcal{T}$- and anti-$\mathcal{T}$-ordering. 
 In the Matsubara representation we  obtain the generating action in the form of  (\ref{z}) after such a symmetrization. Due to   the commutation of photon operators, we   include $-\sfrac{1}{2}$
 in   (\ref{z}).

The photon number moments  $\langle N_{\rm ph}^n \rangle \equiv \langle (\hat\psi^\dagger \hat\psi)^n \rangle$  are given by the derivatives
 \begin{equation}
 \langle N_{\rm ph}^n\rangle=({\rm i})^n\left. \frac{\partial^n}{\partial\xi^n}{ \rm MGF}(\xi)\right|_{\xi=0},
 \end{equation}
while the cumulants   
 are defined as
 \begin{equation}
 	\langle\!\langle 
 	N_{\rm ph}^n\rangle\!\rangle=({\rm i})^n\left. \frac{\partial^n}{\partial\xi^n}{\rm CGF}(\xi)\right|_{\xi=0}. \label{c-n}
 \end{equation}
 Path integration in (\ref{z}) is reduced to the infinite product of Matsubara Green functions involving the counting variable
  \begin{multline}
 	{\rm MGF}(\xi)= e^{{\rm i}\xi /2} \ \frac{\int\limits_0^\infty  e^{ -(\alpha+{\rm i}\xi)\Phi-\gamma \Phi^2}  d\Phi}{\int\limits_0^\infty  e^{ - \alpha \Phi-\gamma \Phi^2} d\Phi}  \prod\limits_{n\neq 0} 
 	\frac{ G_{  n  }(\xi) }{ G_{n }(0)} . \label{mgf-0}
 \end{multline}
The Green function with the counting variable reads as
\begin{equation}
 G_{  n  }(\xi)=\frac{1}{2\pi {\rm i}n  -(\omega+{\rm i}\xi T) - \Sigma_{n}[0]}\ , \ n\neq 0 \ . \label{g-xi}
\end{equation}
Calculation of the integrals  and product in (\ref{mgf-0})  yields for the resonant case  ($\epsilon_j=\bar\epsilon=\omega$):
\begin{equation}
{\rm MGF}(\xi)={\rm MGF}_0(\xi) {\rm MGF}_{\rm fl}(\xi) , \label{mgf} 
\end{equation}
where the zero mode's and quasiparticles' parts are 
 \begin{equation}
 	{\rm MGF}_0(\xi)= \exp\Big[\frac{2{\rm i}\alpha\xi-\xi^2}{4\gamma}  \Big] \frac{ { \rm erfc} \frac{\alpha+{\rm i}\xi}{2\sqrt{\gamma}}}{{ \rm erfc} \frac{\alpha}{2\sqrt{\gamma}}}
\end{equation}
and 	
 \begin{multline}	
	 {\rm MGF}_{\rm fl}(\xi)=\\=\Big[1+\frac{{\rm i}\xi T\omega}{\omega^2-\Omega_T^2}\Big]\frac{(\cosh\frac{\omega}{T}-\cosh\frac{\Omega_T}{T})e^{{\rm i}\xi /2}}{\cosh\!\Big[\!\frac{\omega}{T} +\frac{{\rm i}\xi}{2}\!\Big]-\cosh\sqrt{\frac{\Omega_T^2}{T^2}-\frac{\xi^2}{4}}} \ . 
	\end{multline}
With the use of this result for MGF one can obtain the above expressions for the  photon number and its fluctuations (\ref{n-ph}) and (\ref{c2-ph}).

\subsection{FCS at the phase transition}
 In the thermodynamic limit of large enough $N$, 
 the  leading contribution to  cumulants  is described by that of the zero mode   ${\rm MGF}_0(\xi)$.
 Thus, the  CGF for the critical point is
   \begin{equation}
 	{\rm CGF}_0(\xi)=\frac{ -\xi^2}{4\gamma}+\ln \Big[{ \rm erfc} \frac{ {\rm i}\xi}{2\sqrt{\gamma}} \Big].
 \end{equation}
The first six  cumulants, which follows from ${\rm CGF}_0(\xi)$, are:
\begin{eqnarray}
\langle 
N_{\rm ph}  \rangle&=&\frac{1}{\sqrt{\pi \gamma}}, \\  \langle\!\langle 
N_{\rm ph}^2\rangle\!\rangle&=&\frac{\pi-2}{2\pi \gamma}, \\ \langle\!\langle 
N_{\rm ph}^3\rangle\!\rangle&=&\frac{4-\pi}{2(\pi \gamma)^{3/2}}, \\ \langle\!\langle 
N_{\rm ph}^4\rangle\!\rangle&=&\frac{2(\pi-3)}{(\pi \gamma)^{ 2}}, \\
  \langle\!\langle 
  N_{\rm ph}^5\rangle\!\rangle&=&\frac{96-40\pi+3\pi^2}{4(\pi \gamma)^{5/2}}, \\ \langle\!\langle 
  N_{\rm ph}^6\rangle\!\rangle&=&\frac{60(\pi-2)-7\pi^2}{(\pi \gamma)^{3}}.
\end{eqnarray}
From a numerical calculation it follows that  higher cumulants   alter  their signs, for instance, as it seen from the negativity of  the 5th and 6th ones.
The non-zero cumulants for $n>2$ is the consequence of that fact that photons' probability distribution function   is half of a Gaussian because of  the positively defined variable of integration $\Phi$ in (\ref{mgf-0}).

The  Fourier transformation of the MGF  provides the probability density   to measure  $N_{\rm ph}$ photons on average
\begin{equation}
	\mathcal{P}(N_{\rm ph})= \int\limits_{-\infty}^{\infty} {\rm MGF}(\xi) e^{{\rm i} \xi N_{\rm ph}} d\xi. \label{p}
\end{equation}
Note, that $\mathcal{P}$ is a non-zero function of the continuous variable  $N_{\rm ph}$. This is due to that $N_{\rm ph}$ is not an eigenvalue of    the Hamiltonian (\ref{h-rwa}). Hence,  non-integer values $N_{\rm ph}$ are assumed to be  observed as the  thermodynamical averages.

As long as the $\psi_n$-fluctuations   are frozen out if the system is near  the critical point and $N$ is large enough, one finds from (\ref{z}) and (\ref{p}) that  the probability density  is identical to the exponent in $Z$ (\ref{z}) as $$
\mathcal{P}_0(N_{\rm ph})=2\pi \theta(N_{\rm ph})\frac{\exp[-\alpha N_{\rm ph}-\gamma N_{\rm ph}^2]}{Z(0)}.
$$
In particular,  at the critical point  ${\rm MGF}_0(\xi)$ from (\ref{mgf}) the distribution is
\begin{equation}
\mathcal{P}_{\rm c}(N_{\rm ph})=\begin{cases} 4 \sqrt{\pi\gamma}\exp[-\gamma N_{\rm ph}^2], & \mbox{ if } N_{\rm ph}\geq 0, \\
		0, & \mbox{ if } N_{\rm ph}<0.  
		\end{cases}
\end{equation}
  This is the  half of the Gaussian for $N_{\rm ph}>0$, while  for unphysical  $N_{\rm ph}<0$ it is zero.  
  At  the critical point (we assume below that $\Omega_{\rm c}=\omega$), the distribution's maximum is located at $N_{\rm ph}=0$. In the superradiant phase,  the maximum of $\mathcal{P}(N_{\rm ph})$ is shifted to a non-zero value. In other words, for higher values $\Omega\gg\omega$ one obtains from $\ln [{\rm MGF}_0(\xi)]$ that in  the leading order $\langle N_{\rm ph}\rangle=\frac{N\omega^2}{2\Omega^2}$ and $\langle\!\langle N_{\rm ph}^2\rangle\!\rangle=\frac{NT\omega^3}{2\Omega^4}$. The higher cumulants are strongly suppressed by the  exponent: for instance, the third one is $\langle\!\langle N_{\rm ph}^3\rangle\!\rangle\sim e^{-N\frac{\omega}{T}}$.

\subsection{FCS for weak interaction and normal phase}
\label{seq:normalphase}

In this part we discuss MGF at the normal phase and weak coupling limit.
  It is assumed that the system is   far away from the fluctuational region, i.e., $\Omega_T\ll \omega $ 
  (see Eq. \ref{fluct-zone-normal}). Taking the limit $\gamma\to 0$  in (\ref{mgf}) one obtains  the MGF for the  normal phase of the Dicke model:  
\begin{equation}
{\rm MGF}(\xi)=\frac{(\cosh\frac{\omega}{T}-\cosh\frac{\Omega_T}{T})e^{{\rm i}\xi /2}}{\cosh\!\Big[\!\frac{\omega}{T} +\frac{{\rm i}\xi}{2}\!\Big]-\cosh\sqrt{\frac{\Omega_T^2}{T^2}-\frac{\xi^2}{4}}} .
\end{equation}
In the decoupled limit, where the Rabi frequency is the smallest scale $\Omega_T\ll \{T, \omega\}$, one arrives at the MGF of the free photon mode of the frequency $\omega$ 
\begin{equation}
	{\rm MGF}(\xi
	)=\frac{1-e^{-\beta \omega }}{1-e^{-{\rm i}\xi-\beta\omega}}.  \label{MGF-decoupled}
\end{equation}
Note, that it is $2\pi$-periodic function of the counting variable. The discrete Fourier transformation of (\ref{MGF-decoupled}) at the finite interval $[0;2\pi]$ of the single period
yields  the standard  Hibbs distribution probabilities 
\begin{equation}
	P_{n}=(1-e^{-\beta\omega})e^{-n\beta\omega}, \quad n\geq 0.
\end{equation}
Obviously, the infinite integral definition (\ref{p}) one would  obtains delta-peaks in the probability distribution density located at $N_{\rm ph}=n\geq 0$, being the  eigenvalues of the free photon mode Hamiltonian, as $$\mathcal{P}(N_{\rm ph} )=\frac{1}{2\pi}\sum_{n\geq 0}P_n \delta(N_{\rm ph}-n).$$
Note that the cumulant generating function for the free mode is
\begin{equation}
{\rm CGF}(\xi )={\rm i}\frac{\xi}{2}-\ln \frac{\sinh\frac{\omega +{\rm i}\xi T}{2T}}{\sinh\frac{\omega  }{2T}} \ .  \label{CGF-rwa-0}
\end{equation}
The cumulants itself are
\begin{equation} 
	\langle\!\langle N_{\rm ph}^n\rangle\!\rangle = \begin{cases}
		\frac{1}{2}\coth\frac{\omega  }{2T} -\frac{1}{2}, & n=1; \\ \\
		\frac{(-1)^{n-1}}{2^{n}}\left.\frac{\partial^{n-1} }{\partial x^{n-1}} \coth x \right|_{x=\frac{\omega  }{2T}} , & n\geq 2 .
	\end{cases} 
\end{equation}
One arrives at the mentioned above  Fano factor $
F_0=(1-e^{-\beta \omega })^{-1}
$  in (\ref{F-0}) and the relative fluctuations parameter $r_0=e^{ \beta \omega }$.

\section{Conclusions}\label{sec:concl}
In this work we addressed to fluctuations near  superradiant transition which is driven by an interaction between a single-mode photons and multi-qubit environment.
In such  consideration the collective Rabi frequency is varied (it can be close to the critical value of superradiant transition), while the temperature $T $ is kept unchanged.  We did not assume the thermodynamic limit of infinite qubits number $N$ and consider it as large enough but finite value. Our analysis was focused on two types of competing fluctuations -- the thermal one  and that of the superradiant order parameter. This regime is opposite to the transition by the temperature  studied in Ref.~\cite{popov1988functional}.

 We  used   Majorana   fermion representation of qubits' Pauli operators  in order to formulate a path integral approach.
 Having started  from the Dicke Hamiltonian, we demonstrate how one can derive the effective action for the photon mode,  obtained by alternative fermionization techniques in Refs.~\cite{popov1988functional,eastham2006finite}.  After that we calculated the average photons number and equilibrium  fluctuations  in terms of the effective action formalism.  
As a generalization,  the full counting statistics, providing higher order cumulants of the   photon numbers, was formulated. 

Most of the  result of this paper address a  low temperature   regime and a resonance between qubits and photon mode frequency $\omega$. It was shown that  the  Gaussian approximation for thermal fluctuations is  exact and analytical solution can be found, if $\hbar \omega \gg  k_{\rm B} T \gg  \hbar \omega/N$.   In this limit the critical value of the collective Rabi frequency  is $\Omega_{\rm c}=\omega$ and the average photon number  at this point is  $\langle N_{\rm ph}\rangle = \sqrt{N k_{\rm B} T/(\pi\hbar\omega)}$. The relative fluctuations parameter $r_{\rm c}\equiv\langle\!\langle N_{\rm ph}^2 \rangle\!\rangle/ \langle N_{\rm ph} \rangle^2$, where the second cumulant is $\langle\!\langle N_{\rm ph}^2 \rangle\!\rangle=\langle N_{\rm ph}^2 \rangle-\langle N_{\rm ph} \rangle^2$,
is universal at the critical point $r_{\rm c} =\pi/2-1$. 
A domain  near $\Omega_{\rm c}$ in the superradiant phase, where $r$ is not suppressed, corresponds to the  fluctuational  Ginzburg-Levanyuk region. The width   of such frequency range is proportional to  $\sqrt{\omega k_{\rm B} T/(\hbar N)}$; this is much smaller than $k_{\rm B}T$ and shrinks at thermodynamic limit.  Another characteristic,  Fano factor $F\equiv\langle\!\langle N_{\rm ph}^2 \rangle\!\rangle/ \langle N_{\rm ph} \rangle$, decreases from the unity in   decoupled limit $\Omega\ll\Omega_{\rm c}$ to a minimum $F_{\rm min}<1$ at $\Omega\lesssim\Omega_{\rm c}$. The latter indicates   a negative correlation between photons. The further increase of $\Omega$ up to the critical value results in a significant  growth of the Fano factor to a maximum $F_{\rm c}\approx  \langle N_{\rm ph}\rangle \gg 1$.    This means significantly positive photon-photon correlations at the superradiant transition. There is a reentrance no negative correlations in the superradiant phase as it follows from the  decaying of the Fano-factor above the critical point. 

As a generalization, for opposite  limit of wide spectral distribution of qubits environment  we find $\langle N_{\rm ph}\rangle{\sim} \sqrt{k_{\rm B}T/\delta\epsilon} \ln\frac{\epsilon_{\rm c}}{\omega}$, where $\delta\epsilon$ and $\epsilon_{\rm c}$ are the average level spacing and upper cut-off energy of the spectrum, respectively.  
  For high temperatures, $k_{\rm B} T\gg \hbar \omega$, the neglecting of non-Gaussian fluctuations of quasiparticles is valid for any $N$ --  in contrast to the low-temperature regime. The finiteness of the qubit number can change a behavior of fluctuations at the critical point. Namely, for $\sqrt{k_{\rm B} T/(\hbar\omega)} \gtrsim N $ the quasiparticle fluctuations become greater than that of superradiant order parameter.  This intermediate region shows a non-universal enhancement of $r_{\rm c}$ which reveals a two-level nature  of qubits environment.

We believe that the above results can be of an interest in the context of  state-of-the-art   hybrid systems and quantum metamaterials  operating in   GHz frequency domain. 
The coupling constants $g$ in superconducting systems range from    MHz to several GHz showing  a realization of an ultra-strong coupling regime. The ratios  
of $ g/\omega\sim 0.071$~\cite{Bosman2017},  $g/\omega\sim 0.6$~\cite{Andersen_2017,braumuller2017analog} and $g/\omega \sim 0.72 - 1.34$~\cite{Fumiki2016} have been demonstrated. Consequently,   the critical  qubit number  $N_{\rm c}=(\omega/g)^2$  needed for turning on the superradiant transition can be around $10^0$ to $10^2$. Another possibility for realization of the phase transition are hybrid systems with NV centers in diamonds.  Our estimations are based on  Ref.~\cite{Putz2014}  where individual coupling constant $g\sim 10$ Hz and the number of NV centers   $N\sim 10^{12}$. The collective Rabi frequency $\Omega\sim 20$ MHz is two orders less than the critical value $\Omega_{\rm c}\sim 2$ GHz and, according to our consideration, the system is in the normal phase.   For the above value of $g$, the number $N$ should be increased by four orders  up to the critical  $N_{\rm c}\sim 10^{16}$ in order to reach the superradiant phase.

\section{Acknowledgments}\label{sec:ackn}
The research was funded by the Russian Science Foundation under Grant No. 16-12-00095. 
Authors thank   Andrey A. Elistratov  for fruitful discussions.

\section{Appendix}

 \label{app-corr}

In this Appendix we analyze a role of  the leading non-Gaussian correction   $\propto  \Phi\bar\psi_n\psi_n $, which originates from $\Phi$-dependent self-energies $\Sigma_n[ \Phi ]$ and $\tilde\Sigma_n[ \Phi,\varphi ]$. The analysis  is based on the low-energy effective action  which depends on variables $\Phi$ and $\varphi$. 
Such an action follows from  Gaussian integration in  $S_{\rm eff}[\Phi,\varphi,\bar\psi_n,\psi_n]$ (\ref{s_eff}) over all fluctuations related to  fields $\bar\psi_n,\psi_n$ with $n\neq 0$:
 \begin{equation}
S_{\rm zm}[\Phi,\varphi]=\beta \omega \Phi+ S_{\mathcal{G}}[\Phi,\varphi]+\delta S[\Phi,\varphi] +\ln Z_{\rm ph} \ . \label{app:s-0}
 \end{equation}
The correction $\delta S[\Phi,\varphi]$ here  is the result of the integration;  in the most general form it reads as
 \begin{multline}
	\delta S[\Phi,\varphi]=\\
 \frac{1}{2}\sum\limits_{n\neq 0} {\rm tr}\ln  \beta \!
	\begin{bmatrix}
		-{\rm i}2\pi n T{+}\omega{+}\Sigma_n[\Phi] &  \tilde\Sigma_n[\Phi,\varphi] \\  \\
		(\tilde\Sigma_{-n}[\Phi,\varphi] )^* &   {\rm i}2\pi n T{+}\omega{+}\tilde\Sigma_{-n}[\Phi]
	\end{bmatrix} \ .
	 \label{app:s}
\end{multline}
In what follows we analyze a contribution to the action $\delta S[\Phi,\varphi]$ due to the dependency of the self-energies on $\Phi$.	The normal and anomalous parts of the self-energies originates from the  matrix structure of $S_{\rm fl}$  introduced in (\ref{s-fl}). Their explicit expressions are:
 \begin{widetext}
  \begin{multline}
 \Sigma_n[\Phi] = 
  \Omega^2  \frac{(\omega_n^2+2 i \epsilon \omega_n )   \tanh  \frac{\epsilon }{2 T} }{4 (i \omega_n-\epsilon )   \left(4g^2 \Phi +\omega_n (\omega_n+2 i \epsilon )\right)}      -   \\
  - \Omega^2 \frac{     24 g^4  \Phi^2+2g^2   \Phi  \left(3 \omega_n^2+4 \epsilon ^2+4 i \omega_n \epsilon \right)+i \omega_n \epsilon  \left(\omega_n^2+2 \epsilon ^2+i \omega_n \epsilon \right)   }{4   \sqrt{4g^2 \Phi +\epsilon ^2} \left(4g^2 \Phi +\omega_n^2+\epsilon ^2\right) \left(4g^2 \Phi +\omega_n (\omega_n+2 i \epsilon )\right)} \tanh  \frac{\sqrt{4g^2 \Phi +\epsilon ^2}}{2 T} \ , \label{sigma-normal}  
 \end{multline}	  
 \begin{equation}
\tilde\Sigma_n[\Phi,\varphi] = 
-\frac{\Omega^4 \Phi e^{2{\rm i}\varphi} }{2 \sqrt{4 g^2  \Phi +\epsilon ^2} \left(4 g^2  \Phi +\omega_n^2+\epsilon ^2\right)}\tanh  \frac{\sqrt{4 g^2  \Phi +\epsilon ^2}}{2 T} \ , \ \omega_n=2\pi n T \ . \label{sigma-anomal}
\end{equation} 
	\end{widetext}
We focus below on the fluctuational region near the superradiance which is  second order phase transition.  It means that the   leading contribution  from quantum dynamics of $\Phi$ belongs to a certain region near $\Phi=0$. It follows from (\ref{sigma-normal}) and (\ref{sigma-anomal}) that near $\Phi =0$ the leading $\Phi$-dependent contributions in self-energies are  $\Sigma_n[ \Phi ]\approx\Sigma_n[ 0 ]+\Phi\Sigma'_n[ 0 ] $ and $\tilde\Sigma_n[\Phi,\varphi]\propto \Phi$. We analyze these $\Phi$-dependent parts as perturbations.  It is required  that such perturbations give small contributions to unperturbed part of the  effective action (\ref{app:s-0}). 
Similarly to the previous studies, we limit ourselves by the second order expansion in the unperturbed part: $$\beta \omega \Phi+ S_{\mathcal{G}}[\Phi,\varphi]\approx \alpha \Phi+\gamma\Phi^2 \ .$$
Note that $\alpha$ can be arbitrary small at the critical point while $\gamma$ is always finite. It means that  the linear by $\Phi$ part form $\delta S[\Phi,\varphi]$ is  relevant while higher order terms are not. 
The expansion  of the logarithm in $\delta S[\Phi,\varphi]$ up to the first order by $\Phi$  gives
\begin{equation}
 \delta S[\Phi]=\delta\alpha \ \Phi \ ,  
 \label{app:delta-s-correction}
\end{equation}
where
 \begin{equation}
 \delta\alpha=\sum\limits_{n\neq 0}\frac{\Sigma'_n[0]}{-{\rm i}2\pi n  +\beta\omega+\beta \Sigma_n[ 0 ]} \ . \label{app:delta-alpha}
\end{equation}
In other words, the leading perturbation from $\delta S[\Phi,\varphi]$  is contributed  by the linear part in the normal self-energy $\Sigma_n[ \Phi ]$. 
The anomalous part results in second order corrections which provides small corrections to $\gamma$ and  are not relevant.  The phase $\varphi$ does not appear in this consideration.

We calculate $\delta \alpha$ from (\ref{app:delta-alpha})  for the case of full resonance   $
 \epsilon_j=\bar\epsilon=\omega
 $ and absence of disorder in coupling terms, i.e.  $g_j=g$. 
At the critical point under consideration we set $\Omega=\omega\sqrt{\coth \frac{\omega}{2T}}$ and the result   is 
\begin{multline}
	\delta \alpha_{\rm c}= \frac{\coth  \frac{\beta\omega }{2} }{4 N} \left(6+\frac{\beta^2\omega^2}{1+\cosh\beta\omega}-3 \beta\omega  \coth \frac{\beta\omega }{2}\right) . \label{app:correction}
\end{multline}
The expression (\ref{app:correction}) has the following asymptotics for low temperatures   
\begin{equation}\delta\alpha_{\rm c}(T\ll\omega)=-\frac{3 \omega}{4  N T}.
\end{equation}
 Let us estimate typical value  of $\Phi'$ where the integral over $\exp [-(\alpha+\delta\alpha) \Phi-\gamma\Phi^2]$ in the  partition function does   converge. In the vicinity of the phase transition ($\alpha=0$) it is given by the Gaussian integrand's width, i.e., $\Phi'\sim \gamma^{-1/2}\sim \sqrt{NT/\omega}$.  The requirement that the perturbation   $\delta S[\Phi]$ is small means that it is much less than the unity at the convergence region, namely, $|\delta S[\Phi']|=|\delta\alpha_{\rm c} \ \Phi' |\ll1$.  
The latter results in the  following  condition for low temperatures 
\begin{equation}
\omega\gg T \gg \frac{\omega}{N} \ . \label{app:condition}
\end{equation}
It defines the range of parameters where our approach based on the Gaussian approximation in  $S_{\rm fl}$  (\ref{s-fl-main}) for    fluctuations   is valid.

 It is important that this calculation  provides the small parameter of this theory $\kappa \equiv |\delta S[\Phi']|$. At the critical point of the superradiant transition it is 
 \begin{equation}
 	\kappa_{\rm c}=\sqrt{\frac{  \omega}{   N T}}\ll 1 . \label{app:small-param}
 \end{equation}

Note, that for much lower  temperatures $T\ll  \omega/N $ the non-Gaussian contributions  by $\psi_n$  as well as $\Psi_0$-dependencies in $\Sigma_n$ and $\tilde\Sigma_n$  can not be neglected in  $S_{\rm fl}$. Technically, it means that  the higher order terms in the expansion of (\ref{tr_log})  by $\mathbf{V}[\Psi_\tau]\mathbf{G}_{\tau-\tau'}\mathbf{V}[\Psi_{\tau'}]$ 
should be taken into account.

In the high temperature limit, $T\gg\omega$, the critical coupling is enhanced as $\Omega_{\rm c}=\sqrt{T\omega}$, the correction and convergence region are 
\begin{equation}
\delta\alpha_{\rm c}(T\gg\omega)=-\frac{7}{120N}\left(\frac{ \omega}{T}\right)^3 \ , \quad \Phi'\sim   \sqrt{N}\frac{T}{\omega} \ .
\label{}
\end{equation}
We obtain the following value of the non-Gaussian correction  $|\delta S[\Phi']|=|\delta\alpha_{\rm c} \ \Phi' |\sim  \beta^2\omega^2/\sqrt N$. The value of $|\delta S[\Phi']| $ is small compared to unity for any $N$, in contrast to low temperature limit (\ref{app:condition}).

\end{document}